\documentclass[a4paper,11pt,reqno]{article}
\usepackage{a4wide}
\usepackage{amsmath,amsfonts,amssymb}
\usepackage[english]{babel}
\usepackage{soul}
\usepackage{nicefrac}
\usepackage[mathscr]{euscript}
\usepackage{setspace}
\usepackage{datetime}
\usepackage[nosort]{cite}
\usepackage{mathrsfs}
\usepackage[T1]{fontenc}
\usepackage{txfonts}
\usepackage{kpfonts}
\usepackage{upgreek}
\usepackage{comment}
\usepackage{mathpazo}
\usepackage{graphicx}
\usepackage[breaklinks=true]{hyperref}
\newcommand{\beq}{\begin{equation}}
\newcommand{\eeq}{\end{equation}}
\newcommand{\beqn}{\begin{eqnarray}}
\newcommand{\eeqn}{\end{eqnarray}}
\newcommand{\pa}{\partial}
\numberwithin{equation}{section}

\hypersetup{colorlinks=true,
			urlcolor=blue,
			citecolor=magenta,
			linkcolor=blue,}

\let\oldsqrt\sqrt
\def\sqrt{\mathpalette\DHLhksqrt}
\def\DHLhksqrt#1#2{%
\setbox0=\hbox{$#1\oldsqrt{#2\,}$}\dimen0=\ht0
\advance\dimen0-0.2\ht0
\setbox2=\hbox{\vrule height\ht0 depth -\dimen0}%
{\box0\lower0.4pt\box2}}

\newcommand{\RNum}[1]{\uppercase\expandafter{\romannumeral #1\relax}}

\author{
  \begin{minipage}{.97\linewidth}
    \vspace{1cm}
       \begin{center}
      \begin{small}
               \textbf{Andrea Campoleoni},$^{1,2}$  
               \textbf{Luca Ciambelli},$^3$ 
             \textbf{Charles Marteau},$^3$ 
                     \\
     \textbf{P. Marios Petropoulos}$^3$ and 
      \textbf{Konstantinos Siampos}$^4$
              \end{small}
    \end{center}
    \vspace{0.5cm}
      \hspace{2.4cm}\begin{minipage}{.7\linewidth}
\begin{center}     {\it \begin{footnotesize}
\hbox{
\kern-1.8cm
\vbox{
\begin{itemize}
 \item[$^1$]Service de Physique de l'Univers\\ 
  Champs et Gravitation\\
Universit\'e de Mons -- UMONS\\ 
Place du Parc, 20 \\
7000 Mons, Belgique
      \end{itemize}
      }   
\kern-3cm
      \vbox{
 \begin{itemize}
                  \item[$^2$]
Institut f\"ur Theoretische Physik\\
ETH Z\"urich\\ 
Wolfgang-Pauli-Strasse 27\\ 
8093 Z\"urich, Switzerland               
      \end{itemize}}
    }
     \end{footnotesize}}
\end{center}
    \end{minipage}
    \vspace{0.5cm}\begin{minipage}{.7\linewidth}
\begin{center}     
{\it \begin{footnotesize}
\hbox{\kern0.6cm\vbox{\vskip0cm
 \begin{itemize}
              \item[$^3$]             CPHT -- Centre de Physique Th\'eorique\\ 
        Ecole Polytechnique, CNRS UMR 7644\\
        Universit\'e Paris--Saclay\\
        91128 Palaiseau Cedex, France
                            \end{itemize}}
\kern-3cm\vbox{
\begin{itemize}
 \item[$^4$]       Theoretical Physics Department\\
CERN\\ 
1211 Geneva 23, Switzerland
\vskip0.45cm 
       \end{itemize}\vskip0.05cm
}
}
     \end{footnotesize}}
\end{center}
     \end{minipage}
  \end{minipage}
}

\title{\vspace{1.5cm}
 \boldmath 
    \textbf{Two-dimensional fluids and their holographic duals}
  \unboldmath
}

\date{}

\begin{document}

\begin{titlepage}
\maketitle
\thispagestyle{empty}

 \vspace{-14.cm}
  \begin{flushright}
  CPHT-RR078.082018\\
CERN-TH-2018-231
  \end{flushright}
 \vspace{11.7cm}

\begin{center}
\textsc{Abstract}\\  
\vspace{0.5cm}	
\begin{minipage}{1.0\linewidth}

We describe the dynamics of two-dimensional relativistic and Carrollian fluids. These are mapped holographically to three-dimensional locally anti-de Sitter and locally Minkowski spacetimes, respectively. To this end, we use Eddington--Finkelstein coordinates, and grant general curved two-dimensional geometries as hosts for hydrodynamics. This requires to handle the conformal anomaly, and the expressions obtained for the reconstructed bulk metrics incorporate non-conformal-fluid data. We also analyze the freedom of choosing arbitrarily the hydrodynamic frame for the description of relativistic fluids, and propose an invariant entropy current compatible with classical and extended irreversible thermodynamics.  This local freedom breaks down in the dual gravitational picture, and fluid/gravity correspondence turns out to be sensitive to dissipation processes: the fluid heat current is a necessary  ingredient for reconstructing all Ba\~nados asymptotically anti-de Sitter solutions. The same feature emerges for Carrollian fluids, which enjoy a residual frame invariance, and their Barnich--Troessaert locally Minkowski duals.  These statements are proven by computing the algebra of surface conserved charges in the fluid-reconstructed bulk three-dimensional spacetimes.

\end{minipage}
\end{center}


\end{titlepage}

\onehalfspace

\begingroup
\hypersetup{linkcolor=black}
\tableofcontents
\endgroup
\noindent\rule{\textwidth}{0.6pt}

\section{Introduction}

Fluid/gravity correspondence is a macroscopic spin-off of holography, originally mapping relativistic fluid configurations onto Einstein spacetimes,  \emph{i.e.}
spacetimes whose Ricci tensor is proportional to the metric. These are obtained in the form of a derivative expansion \cite{Bhattacharyya:2007, Haack:2008cp, Bhattacharyya:2008jc, Hubeny:2011hd}, inspired from the fluid homonymous expansion (see \emph{e.g.} \cite{Kovtun:2012rj, Romatschke:2009im}).
An alternative reconstruction of Einstein spacetimes from boundary data is based on the Fefferman--Graham theorem \cite{PMP-FG1, PMP-FG2}, which provides an expansion in powers of a radial space-like coordinate in the so-called Fefferman--Graham gauge.

Compared to the radial Fefferman--Graham expansion, the derivative expansion 
has several distinctive features listed hereafter.
\begin{itemize}
\item The boundary data in the Fefferman--Graham expansion are the first and second fundamental forms, interpreted as the boundary metric and the boundary fluid energy--momentum tensor. For the derivative expansion, the boundary data include also a vector congruence, whose derivatives set the order of the expansion. This congruence is interpreted as the boundary fluid velocity field.
\item The derivative expansion is not built along a spatial but rather a null radial coordinate, whose differential form is the dual of the fluid velocity vector. It  is implemented in Eddington--Finkelstein coordinates, and provides radial fall-offs which are slightly less restrictive than those of the Bondi gauge \cite{Bondi1962, Sachs1962}.
\item The derivative expansion is well behaved in the Ricci-flat limit (vanishing bulk scalar curvature, \emph{i.e.} cosmological constant).
\end{itemize}
The last property has recently allowed to set up a derivative expansion for asymptotically flat spacetimes, establishing thereby, at least macroscopically, a holographic correspondence among Ricci-flat bulk solutions and boundary Carrollian hydrodynamics \cite{CMPPS2}, which is the ultra-relativistic (vanishing velocity of light) limit of fluid dynamics. The derivative expansion in Eddington--Finkelstein coordinates has been instrumental in reaching this result, because the Fefferman--Graham expansion is ill-defined in the limit of vanishing cosmological constant.

The first of the above three features raises another important question, regarding the role played by the boundary fluid congruence. 
In this respect, we remind that the velocity field of a relativistic fluid can be chosen freely, altering neither the energy--momentum tensor nor the entropy current, but only transforming the various pieces that enter the decomposition of these quantities with respect to its longitudinal and transverse directions \cite{Landau}. This is usually referred to as the hydrodynamic-frame invariance.

The fluid congruence appears explicitly in the derivative expansion, as we will discuss in the following. 
Conforming to the above fluid-dynamics logic, one could consider another fluid frame. This would leave the boundary metric and energy--momentum tensor unchanged, and the corresponding reconstructed bulk metric would be amenable to its former expression by an appropriate bulk diffeomorphism. Still, this diffeomorphism might be large, in which case the two boundary hydrodynamic frames would lead to definitely distinct dual spacetimes with different global properties.

Analyzing the role of the velocity field in the fluid/gravity derivative expansion is not an easy task. Generically this derivative expansion
is organized in the form of a series, whose order is set by the derivatives of the velocity field, and which is designed to comply with Weyl covariance. Furthermore, in the original works  \cite{Bhattacharyya:2007, Haack:2008cp, Bhattacharyya:2008jc, Hubeny:2011hd}, this series was  expressed using a specific  hydrodynamic frame known as Landau--Lifshitz. In this context it is difficult to investigate the global behaviour under a congruence transformation, since typically only the first few orders in the expansion are available. In some more specific classes, it is possible to resum the derivative expansion (see \cite{Caldarelli:2012cm, Mukhopadhyay:2013gja, Petropoulos:2014yaa, Gath:2015nxa, Petropoulos:2015fba}), which could help circumventing the latter difficulty. In order to resum the expansion, one   needs to abandon the Landau--Lifshitz frame, and impose integrability conditions relating the heat current and stress tensor (\emph{i.e.} the non-perfect components of the energy--momentum tensor) to the boundary geometry. The integrability conditions, however, are not covariant under changes of fluid congruence. Hence, the benefit of adopting resummed expressions is tempered when coming to the point of hydrodynamic-frame transformations.

Substantial simplifications occur in three bulk dimensions. On the one hand, all expansions, Fefferman--Graham or derivative, are naturally truncated to a finite number of terms. 
On the other hand, asymptotically anti-de Sitter spacetimes are locally anti-de Sitter. As a consequence the distinction among Einstein solutions is exclusively encoded in their global properties,  labeled unambiguously by their conserved surface charges, as \emph{e.g.} in Ba\~nados solutions \cite{Banados_solutions}. Probing the fluid/gravity hydrodynamic-frame invariance amounts therefore to analyze the conserved charges and their algebra in different fluid frames. This is one of the aims of the present work, and we will show that contrary to the naive expectation,\footnote{The question of global versus local properties of bulk solutions in relation with the dual boundary fluid was mentioned in
the Appendix B of Ref.  \cite{Bhattacharyya:2008jc}. This discussion is not conclusive though, in particular because of the absence of any charge computation, which would have allowed to make concrete statements about the landscape of locally anti-de Sitter spacetimes and their dual fluids.}  changing fluid frame can alter the global properties of the reconstructed Einstein spacetime.

As already mentioned, the derivative expansion in Eddington--Finkelstein coordinates admits a well-defined limit of vanishing cosmological constant. This limit generalizes the customary fluid/gravity correspondence to a duality between Ricci-flat spacetimes and Carrollian hydrodynamics emerging at null infinity  \cite{CMPPS1}.
In some instances, Carrollian fluids possess a residual frame invariance involving a kinematical parameter reminiscent of the relativistic velocity field. The latter enters the flat derivative expansion, and it is legitimate to ask the same questions about the role of frame invariance as for anti-de Sitter spacetimes.  Again, answering is possible in three dimensions, where the derivative expansion admits a finite number of terms, and all Ricci-flat spaces are locally Minkowskian. These are globally distinguishable by conserved surface charges, as \emph{e.g.} for the family obtained in \cite{Barnich:2010eb} with appropriate fall-off conditions that will be referred to as \emph{Barnich--Troessaert} solutions.

In order to undertake the above analysis we will 
set up the fluid/gravity derivative expansions in three dimensions.\footnote{\emph{Expansion} is an abuse of terminology in three dimensions because there, it is naturally  truncated. We will often make it, and use the word \emph{resummation} for simple sums.} In other words, we will obtain expressions providing the bulk dual (Einstein or Ricci-flat) of an arbitrary fluid, hosted by any two-dimensional geometry. Such expressions were not available in full generality for the relativistic fluids, and were unknown for  Carrollian (\emph{i.e.} ultra-relativistic) fluids.

In the relativistic case, we exhibit a universal resummation formula, which turns out to be a BMS-like  (Bondi--Metzner--Sachs,  \cite{Bondi1962, Sachs1962}) alternative to the existing Fefferman--Graham expression \cite{ Barnich:2010eb,skesol}. The prime virtue of our practice is to accommodate the conformal anomaly arising from the curvature of the boundary, which has been ignored in earlier fluid/gravity literature \cite{Haack:2008cp,Bhattacharyya:2008jc} and has a detectable counterpart in the Carrollian situation. For the latter, our fluid reconstruction of flat spacetimes resembles the general formulas given in BMS gauge in \cite{Barnich:2010eb}.

After having settled the derivative expansions, we express the asymptotic  charges\footnote{Useful references for the analysis of asymptotic charges are \emph{e.g.} \cite{BarnichCG1, BarnichCG2}. Our surface-charge computations have been performed with the package \cite{code}, built using the conventions of the papers just quoted.} of the reconstructed spacetimes in terms of the fluid data and we prove that the choice of frame may affect the global properties of the solutions. Indeed, we show that the holographic reconstruction of all \emph{Ba\~nados} and \emph{Barnich--Troessaert} solutions requires the boundary fluid (relativistic or Carrollian) have a non-vanishing heat current. In this instance, the charge algebra is either Virasoro or BMS with the expected central charges. Setting the heat current to zero, the solutions carry surface charges obeying algebras of the same type, where the central charges can be trivially reabsorbed though. 


In Sec. \ref{sec:fluids} we review two-dimensional relativistic conformal fluid dynamics, and expand its Carrollian limit, insisting on the hydrodynamic-frame invariance. Section \ref{sec:recon} is devoted to the general method of holographic reconstruction of asymptotically AdS and flat spacetimes. This method is applied in Sec. \ref{sec:recon-flat-bry} for flat two-dimensional boundary metrics, without loosing generality, and followed by the computation of charges, which enables us to reach a clear image of the solutions under investigation. 

Before moving to the main part of the paper, we should add that Sec. \ref{sec:relfluids}  includes a part dedicated to the entropy current of relativistic two-dimensional conformal fluids. Contrary to the energy--momentum tensor the entropy current has no general microscopic definition for systems that are only at local thermodynamic equilibrium. It is usually constructed phenomenologically, in a given hydrodynamic frame, order by order in the velocity and temperature derivative expansion, and subject to several physical conditions. We propose here an entropy current, which fulfills all known criteria, has a  closed form that can be expanded in a non-trivial infinite series, and is explicitly hydrodynamic-frame invariant. This last feature is the backbone of fluid frame invariance.

\section{Two-dimensional fluids}\label{sec:fluids}

\subsection{Relativistic fluids}\label{sec:relfluids}

\subsubsection*{General properties}

We consider a two-dimensional geometry $\mathscr{M}$ equipped with a metric $\text{d}s^2=g_{\mu\nu}\text{d}x^\mu\text{d}x^\nu$.  The dynamics of a relativistic fluid is captured by the energy--momentum tensor $\text{T}=T_{\mu \nu}\text{d}x^\mu\text{d}x^\nu$, which is symmetric  ($T_{\mu\nu}=T_{\nu\mu}$) and generally obeys:
\begin{equation}
\label{T-cons}
\nabla^\mu T_{\mu\nu}=f_\nu,
    \end{equation}
where $f_\nu$ is an external force density.
Together with the equation of state (local thermodynamic equilibrium is assumed), this set of equations provide the hydrodynamic equations of motion. 
Normalizing the velocity  congruence $\text{u}$ as $\| \text{u} \|^2=-k^2$ ($k$ plays the role of velocity of light), we can in general decompose the energy--momentum tensor as
\begin{equation}\label{T} 
T_{\mu \nu}=(\varepsilon+p) \frac{u_\mu  u_\nu}{k^2} +p  g_{\mu\nu}+ \tau_{\mu\nu}+\frac{u_\mu  q_\nu}{k^2}+ \frac{u_\nu  q_\mu}{k^2}
\end{equation}
with $p$ the local pressure and  $\varepsilon$ the local energy density:  
\begin{equation}
\label{long} 
\varepsilon=\frac{1}{k^2}T_{\mu \nu} u^\mu u^\nu.
\end{equation}
 The symmetric viscous stress tensor $\tau_{\mu \nu}$ and the heat current $q_\mu$ are purely transverse:
\begin{equation}\label{trans}
u^\mu    \tau_{\mu \nu}=0, \quad u^\mu  q_\mu =0, \quad 
q_\nu= -{\varepsilon} u_\nu-u^\mu  T_{\mu \nu}.
\end{equation}

In two dimensions, the transverse direction with respect to  $\text{u}$ is entirely supported by the 
Hodge-dual $\ast\text{u}$:\footnote{Our conventions are: $ \eta_{\sigma\rho}=\sqrt{g} \epsilon_{\sigma\rho}$ with $\epsilon_{01}=+1$. Hence $\eta^{\mu\sigma}\eta_{\sigma\nu}=\delta^\mu_\nu$.}
\begin{equation}
 \ast u_\rho=u^\sigma \eta_{\sigma\rho}.
\end{equation}
This dual congruence is space-like and normalized as $\| \ast\text{u} \|^2=k^2$.
Therefore
\begin{equation}
\label{chimag}
\text{q}=\chi \ast \text{u} \quad\text{with} \quad \chi=-\frac{1}{k^2}\ast u^\mu T_{\mu\nu}u^\nu,
\end{equation}
the local \emph{heat density}, appearing here as the magnetic dual of the energy density.  Similarly,  the viscous stress tensor has a unique component encoded in the \emph{viscous stress scalar} $\tau$:\footnote{This component of the energy--momentum tensor is also referred to as the \emph{viscous bulk pressure}, or the \emph{dynamic pressure}, or else the \emph{non-equilibrium pressure}.}
\begin{equation}\label{viscstr}
 \tau_{\mu \nu}= \tau  h_{\mu \nu} \quad\text{with} \quad h_{\mu\nu}=\dfrac{1}{k^2}\ast u_{\mu} \ast u_{\nu}
\end{equation}
the projector onto the space transverse to the velocity field. The trace reads: 
$T^\mu_{\hphantom{\mu}\mu}=p-\varepsilon+\tau$.

The pressure $p$ and the viscous stress scalar $\tau$ appear in the fully transverse component of the energy--momentum tensor. Their sum is therefore the total stress. If the system is free and at \emph{global} equilibrium, $\tau$ vanishes and the stress is given by the thermodynamic pressure $p$ alone. Hence, the viscous stress scalar $\tau$
is usually expressed as an expansion in temperature and velocity gradients, and this distinguishes it from  $p$. The same holds for the heat current $\text{q}$. The coefficients of these expansions characterize the transport phenomena occurring in the fluid. 

The shear and the vorticity vanish identically in two spacetime dimensions. The only non-vanishing first-derivative  tensors of the velocity are the acceleration and  the expansion
\begin{equation}
a_\mu =u^\nu \nabla_\nu u_\mu , \quad
\Theta=\nabla_\mu  u^\mu , \label{def21}
\end{equation}
and one defines similarly 
the expansion of the dual congruence as\footnote{The hodge-dual of a scalar is a two-form and would spell with a suffix star. Instead, $\Theta^{\ast}$ is just another scalar.}
\begin{equation}
\Theta^{\ast}=\nabla_\mu  \ast u^\mu,
\end{equation} 
which enables us expressing the acceleration:
\begin{equation}
\label{caracclim} 
a_\mu
=
\Theta^{\ast}
\ast u_\mu.
\end{equation} 
In first-order hydrodynamics\footnote{For any vector $\text{v}$ and a function $f$, $\text{v}(f)$ stands for $v^\mu\partial_\mu f$. We remind the following identities: $\text{d}^\dag\text{d}f=-\Box f$ with $\text{d}^\dag\text{w}=\ast\text{d}\ast\text{w}=-\nabla^\mu w_\mu$ and 
$\text{d}f= \frac{1}{k^2}\left(\ast \text{u}(f)\ast \text{u}- 
\text{u}(f)\text{u}
\right)$, $\ast\text{d}f= \frac{1}{k^2}\left(\ast \text{u}(f)\text{u}-\text{u}(f) \ast 
\text{u}
\right)$.} 
\begin{eqnarray}\label{e1}
\tau_{(1)}&=&-\zeta \Theta,\\
\label{q1} \chi_{(1)}&=& -\dfrac{\kappa}{k^2}\left(\ast \text{u} (T)+T \Theta^{\ast} \right).
\end{eqnarray}
As usual, $\zeta$ is the bulk viscosity and $\kappa$ is the thermal conductivity -- assumed constant in this expression. 

It is convenient to use the orthonormal Cartan frame $\{\nicefrac{\text{u}}{k}, \nicefrac{\ast \text{u}}{k} \}$. Then the metric  reads:
\begin{equation}
\text{d}s^2
=
\frac{1}{k^2}\left(-\text{u}^2+
 \ast \text{u}^2
\right),
\label{ds2gen}
\end{equation}
while the energy--momentum tensor takes the form:
\begin{equation}
\text{T}=\frac{1}{2k^2}\left(\left(\varepsilon+\chi\right)\left(\text{u}+\ast \text{u}\right)^2+\left(\varepsilon-\chi\right)\left(\text{u}-\ast \text{u}\right)^2\right)
+\frac{1}{k^2}(p-\varepsilon+\tau) \ast \text{u}^2.
\label{Tgen}
\end{equation}

In holographic systems, the boundary enjoys remarkable conformal properties as it defines a conformal class, rather than a specific metric. Under Weyl transformations 
\begin{equation}
\label{conmet}
\text{d}s^2\to  \frac{\text{d}s^2}{{\cal B}^2},
\end{equation}
the velocity form components $u_\mu$ are traded for $\nicefrac{u_\mu}{{\cal B}}$, the energy and heat densities have weight $2$, and the local-equilibrium equation of state is conformal
\begin{equation}\label{confluid}
\varepsilon=p,
\end{equation}
which is accompanied by Stefan's law ($\sigma$ is the Stefan--Boltzmann constant):
\begin{equation}
\label{stefan-3}
\varepsilon = \sigma T^2.
\end{equation}  
Hence, the trace of the energy--momentum tensor is $\tau$. In the absence of anomalies it vanishes and $T_{\mu\nu}$ is invariant under \eqref{conmet}. If $\tau$ is  non-vanishing, the fluid is not conformal and $\tau$ is an anomalous weight-$2$ quantity.

Covariantization with respect to rescalings requires to introduce a Weyl connection one-form \cite{Loganayagam:2008is,Diles:2017azi}, see also Appendix D of \cite{Waldbook}:\footnote{The explicit form of $\text{A}$ is obtained by demanding $\mathscr{D}_{\mu}u^{\mu}=0$ and $u^{\lambda}\mathscr{D}_{\lambda}u_{\mu}=0
$.}
\begin{equation}
\label{Wconc}
\text{A}=\frac{1}{k^2}\left(\text{a} -\Theta \text{u}\right)=
\frac{1}{k^2} \left( \Theta^{\ast} \ast\text{u}  -\Theta \text{u}  \right),
\end{equation}
which transforms as $\text{A}\to\text{A}-\text{d}\ln {\cal B}$. Ordinary covariant derivatives $\nabla$ are thus traded for the Weyl covariant combination $\mathscr{D}=\nabla+w\,\text{A}$, $w$ being the conformal weight of the tensor under consideration.  
We provide for concreteness the Weyl covariant derivative of a form $v_\mu$ and of a scalar function $\Phi$, both of weight $w$:
\begin{equation}
\label{Wv}
\begin{split}
&\mathscr{D}_\nu v_\mu=\nabla_\nu v_\mu+(w+1)A_\nu v_\mu + A_\mu v_\nu-g_{\mu\nu} A^\rho v_\rho,\\ 
&\mathscr{D}_\nu \Phi=\partial_\nu \Phi+ w A_\nu\Phi.
\end{split}
\end{equation}

The Weyl covariant derivative is metric-compatible with effective torsion:
\begin{eqnarray}
\mathscr{D}_\rho g_{\mu\nu}&=&0,\\
\left(\mathscr{D}_\mu\mathscr{D}_\nu -\mathscr{D}_\nu\mathscr{D}_\mu\right)  \Phi&=& w  \Phi F_{\mu\nu},
\end{eqnarray}
where  
\begin{equation}
\label{F}
F_{\mu\nu}=\partial_\mu A_\nu-\partial_\nu A_\mu
\end{equation}
is the Weyl-invariant field strength. Its dual
\begin{equation}
\label{scalF}
F=\ast\text{dA}=\eta^{\mu\nu}\partial_\mu A_\nu=\frac{1}{k^2}\left(\ast\text{u}(\Theta)
-\text{u}(\Theta^\ast) \right)
\end{equation}
is a weight-$2$ scalar.

Commuting the Weyl-covariant derivatives acting on vectors, one defines the Weyl covariant Riemann tensor 
\begin{equation}
\left(\mathscr{D}_\mu\mathscr{D}_\nu -\mathscr{D}_\nu\mathscr{D}_\mu\right) V^\rho=
\mathscr{R}^\rho_{\hphantom{\rho}\sigma\mu\nu} V^\sigma+ w  F_{\mu\nu} V^\rho
\end{equation}
($V^\rho$ are weight-$w$)
and the usual subsequent quantities. 
In two spacetime dimensions, the covariant Ricci tensor (weight-$0$) and the scalar (weight-$2$) curvatures read: 
\begin{eqnarray}
\mathscr{R}_{\mu\nu}&=&{R}_{\mu\nu}+ 
g_{\mu\nu}\nabla_\lambda A^\lambda-F_{\mu\nu}
,
\label{curlRic}
\\
\mathscr{R}&=&R +2\nabla_\mu A^\mu.
\label{curlRc}
\end{eqnarray}
It turns out that 
${R}_{\mu\nu}+ 
g_{\mu\nu}\nabla_\lambda A^\lambda$
vanishes identically. 
Hence
\begin{equation}
\label{Wcurv}
\mathscr{R}=0\Leftrightarrow R=2\text{d}^\dagger \text{A}\quad \text{and} \quad\mathscr{R}_{\mu\nu}=-F_{\mu\nu}.
\end{equation}
The ordinary scalar curvature has a weight-2 anomalous transformation
\begin{equation}
\label{anomR}
R\to \mathcal{B}^2\left(R +2 \Box \ln \mathcal{B}\right)
\end{equation}
(the box operator is here referring to the metric before the Weyl transformation).

\subsubsection*{Hydrodynamic equations and the hydrodynamic-frame covariance}

Using the above tools as well as the identity 
\begin{equation}
\nabla^\mu T_{\mu\nu}=\mathscr{D}^\mu T_{\mu\nu}-A_\nu T^\mu_{\hphantom{\mu}\mu},
\end{equation}
(based on Eqs. \eqref{Wv} and Leibniz rule, for a weight-$0$, rank-2 symmetric tensor), 
the general fluid equations \eqref{T-cons} with $\varepsilon = p$, projected on the light-cone directions $\text{u}\pm \ast\text{u}$  read:\footnote{Notice that any congruence with $w=-1$ in two dimensions obeys 
$ \mathscr{D}_\mu u_{\nu}=\nabla_{\mu}u_{\nu}+\frac{1}{k^2}u_{\mu}a_{\nu}-\Theta h_{\mu\nu} =0$
due to the absence of shear and vorticity,
and similarly  $\mathscr{D}_\mu \ast u_{\nu}=0$.} 
\begin{equation}
\label{T-cons-el-mag-nc} 
 \begin{cases}
\left(u^\mu+\ast u^\mu  \right)\mathscr{D}_\mu \left(\varepsilon+
 \chi
\right)+\left(u^\mu-\ast u^\mu\right)f_\mu
=
-\Theta \tau-\Theta^\ast \tau-\ast \text{u}(\tau)
,\\
\left(u^\mu-\ast u^\mu\right)\mathscr{D}_\mu \left(\varepsilon-
 \chi
\right)+\left(u^\mu+\ast u^\mu\right)f_\mu
=-\Theta \tau+\Theta^\ast \tau+\ast \text{u}(\tau).
\end{cases}
\end{equation} 
Equivalently, these equations are expressed as
\begin{equation}
\begin{cases}
\label{T-cons-el-mag-sol}
\text{d}\left(\sqrt{\varepsilon+\chi+\nicefrac{\tau}{2}}(\text{u}+\ast\text{u})\right)+\dfrac{1}{2\sqrt{\varepsilon+\chi+\nicefrac{\tau}{2}}}(\text{u}-\ast\text{u})\wedge\ast \left(\text{f}-\frac{1}{2}\text{d} \tau\right)=0\,,\\
\text{d}\left(\sqrt{\varepsilon-\chi+\nicefrac{\tau}{2}}(\text{u}-\ast\text{u})\right)-\dfrac{1}{2\sqrt{\varepsilon-\chi+\nicefrac{\tau}{2}}}(\text{u}+\ast\text{u})\wedge\ast  \left(\text{f}-\frac{1}{2}\text{d} \tau\right)=0\,.
\end{cases}
\end{equation}  

Changing hydrodynamic frame, \emph{i.e.} the fluid velocity field, amounts to perform an arbitrary local Lorentz transformation on the Cartan mobile frame
\begin{equation}
\label{locLor}
\begin{pmatrix}
\text{u}^\prime\\
\ast \text{u}^\prime\
\end{pmatrix}
=
\begin{pmatrix}
\cosh \psi(x)&\sinh \psi(x) \\
\sinh \psi(x)&\cosh \psi(x)
\end{pmatrix}
\begin{pmatrix}
\text{u}\\
\ast \text{u}
\end{pmatrix},
\end{equation}
or for the null directions $\text{u}^\prime\pm \ast \text{u}^\prime
=\left(\text{u}\pm\ast\text{u}\right)\,\text{e}^{\pm\psi}$.
This affects the Weyl connection and Weyl curvature scalar as follows
\begin{eqnarray}
\text{A}^\prime&=&\text{A}-\ast\text{d}\psi
\\
F^\prime&=&F+\Box \psi
.
\end{eqnarray}

The transformation  \eqref{locLor}
keeps the energy--momentum tensor invariant provided the energy density and the heat density transform appropriately. Imposing that in the new frame \eqref{confluid} holds, \emph{i.e.} 
$\varepsilon^\prime = p^\prime$, we conclude that 
\begin{equation}
\label{locLor-en-he-nc}
\begin{pmatrix}
\varepsilon^\prime\\
\chi^\prime\
\end{pmatrix}
=
\begin{pmatrix} 
\cosh2 \psi(x)&-\sinh 2\psi(x) \\
-\sinh 2\psi(x)&\cosh 2\psi(x)
\end{pmatrix}
\begin{pmatrix}
\varepsilon\\
\chi
\end{pmatrix}
+\tau \sinh \psi(x)
\begin{pmatrix}
 \sinh \psi(x)\\
 - \cosh \psi(x)
\end{pmatrix}, 
\end{equation}
while, due to the invariance of the trace,
\begin{equation}
\label{locLor-tau}
\tau^\prime=\tau.
\end{equation}
Equivalently one can use 
$\sqrt{\left(\varepsilon^\prime\pm\chi^\prime+\frac{\tau^\prime}{2}\right)}=\sqrt{\left(\varepsilon\pm\chi+\frac{\tau}{2}\right)}\,\text{e}^{\mp\psi}$.

The energy--momentum tensor can be diagonalized with a specific local Lorentz transformation. By definition, the corresponding hydrodynamic frame is the Landau--Lifshitz frame, where the heat current $\chi_{\text{LL}}$ is vanishing.  We find 
\begin{equation}
\text{T}= \frac{\varepsilon_{\text{LL}}}{k^2}\text{u}_{\text{LL}}^2 +\frac{\varepsilon_{\text{LL}}+\tau}{k^2}  \ast \text{u}_{\text{LL}}^2
\label{TLLnc}
\end{equation}
since $\tau_{\text{LL}}=\tau$ and $\chi_{\text{LL}}=0$. The latter condition allows to find the local boost towards the Landau--Lifshitz frame
\begin{equation}
\text{e}^{4\psi_{\text{LL}}}=\frac{\varepsilon+\chi+\nicefrac{\tau}{2}}{\varepsilon-\chi+\nicefrac{\tau}{2}}.
\end{equation}
With this, the  eigenvalues are easily computed. One finds the Landau--Lifshitz energy density
\begin{equation}
\varepsilon_{\text{LL}}=\sqrt{\left(\varepsilon+\chi+\frac{\tau}{2}\right)
\left(\varepsilon-\chi+\frac{\tau}{2}\right)}-\frac{\tau}{2}.
\label{eigene}
\end{equation}
It exhibits an upper bound for $\chi^2$, $\chi^2_{\text{max}}=\left(\varepsilon+\nicefrac{\tau}{2}\right)^2$, which translates causality and unitarity properties of the underlying microscopic field theory. 
The eigenvalue\footnote{We make the reasonable assumption that the fluid energy density is positive. This is generically true, although some exceptions exist. One of those is global $\text{AdS}_3$, indeed realized with a negative-energy dual fluid, whereas the conventional zero-energy fluid reconstructs one Poincar\'e patch of $\text{AdS}_3$.} $\varepsilon_{\text{LL}}$ is supported by the time-like eigenvector
 \begin{equation}
\text{u}_{\text{LL}}=\frac{1}{2}
\left(\left(\frac{\varepsilon+\chi+\nicefrac{\tau}{2}}{\varepsilon-\chi+\nicefrac{\tau}{2}}\right)^{\nicefrac{1}{4}}(\text{u}+\ast\text{u})
+
\left(\frac{\varepsilon-\chi+\nicefrac{\tau}{2}}{\varepsilon+\chi+\nicefrac{\tau}{2}}\right)^{\nicefrac{1}{4}}(\text{u}-\ast\text{u})
\right),
\label{eigenU}
\end{equation}
whereas 
\begin{equation}
\varepsilon^{\ast}_{\text{LL}}=\varepsilon_{\text{LL}}+\tau=\sqrt{\left(\varepsilon+\chi+\frac{\tau}{2}\right)
\left(\varepsilon-\chi+\frac{\tau}{2}\right)}+\frac{\tau}{2}
\label{eigenest}
\end{equation}
is the eigenvalue along the space-like eigenvector $\ast \text{u}_{\text{LL}}$. 
Using the above expressions in the Landau--Lifshitz frame, the fluid equations \eqref{T-cons-el-mag-sol} are recast as follows
\begin{equation}
\begin{cases}
\label{T-cons-el-mag-LL}
2\sqrt{\varepsilon_{\text{LL}}}\text{d}^\dag\left(\sqrt{\varepsilon_{\text{LL}}}\text{u}_{\text{LL}}\right)
-\text{u}_{\text{LL}}\cdot\text{f}-\Theta_\text{LL}\tau
=0\,,\\
2\sqrt{\varepsilon^{\ast}_{\text{LL}}}\text{d}^\dag\left(\sqrt{\varepsilon^{\ast}_{\text{LL}}}\ast \text{u}_{\text{LL}}\right)
+\ast \text{u}_{\text{LL}}\cdot \text{f}+\Theta^{\ast}_\text{LL}\tau
=0\,.
\end{cases}
\end{equation}  

A non-anomalous conformal fluid in two dimensions is defined through the relations \eqref{confluid}, \eqref{stefan-3} and
\begin{equation}\label{confluidtau}
\tau =0.
\end{equation}
Under these assumptions, the last term of \eqref{Tgen} drops, whereas following the fluid equations 
\eqref{T-cons-el-mag-sol} at zero external force ($\text{f}=f_\mu\text{d}x^\mu=0$), the forms $\sqrt{\varepsilon\pm\chi}(\text{u}\pm\ast\text{u})$ are \emph{closed}, and 
can be used to define a privileged light-cone coordinate system, adapted to the fluid configuration. In this specific case, the 
on-shell Weyl scalar curvature reads
 \begin{equation}
\label{Wcurv-expl}
F=-\frac12  \Box\ln\sqrt{\frac{\varepsilon+\chi}{\varepsilon-\chi}}.
\end{equation}  
For conformal fluids, the hydrodynamic-frame transformation \eqref{locLor} acts on the energy and heat densities as a spin-two electric--magnetic boost, the energy being electric and the heat  magnetic.

\subsubsection*{The entropy current}

We would like to close this overview on two-dimensional conformal fluids with the entropy current. 
The entropy appears in Gibbs--Duhem equation  
\begin{equation}
\label{thermorel}
Ts=p+\varepsilon,
\end{equation}  
and is easily computed for conformal fluids in terms of the energy density, using Eq. \eqref{confluid} and Stefan's law \eqref{stefan-3}:
\begin{equation}
\label{s-conf}
s=2 \sqrt{\sigma\varepsilon}.
\end{equation}  

The entropy current is an involved concept because, among other reasons, no microscopic definition is available for out-of-global-equilibrium systems. In arbitrary dimension, there is no generic and closed expression in terms of the dissipative  tensors for  this current, which is generally constructed order by order as a derivative expansion (see \cite{RZ}). Whether this expansion can be hydrodynamic-frame invariant, and at the same time compatible with the underlying already quoted microscopic laws (unitarity and causality) as well as with the second law of thermodynamics is not known in full generality, although this is in principle part of the rationale behind frame invariance. 

In two dimensions, the ingredients for building a hydrodynamic-frame-invariant entropy current are the time-like invariant vector  $\text{u}_{\text{LL}}$ (given in \eqref{eigenU}) and its space-like dual $\ast \text{u}_{\text{LL}}$, plus the invariant scalars $\varepsilon_{\text{LL}}$ and  $ \varepsilon^{\ast}_{\text{LL}}$ (or any combination, see \eqref{eigene} and \eqref{eigenest}). The entropy current should have non-negative divergence, vanishing for a free (\emph{i.e.} at zero external force) perfect fluid. In the case at hand, a perfect fluid is necessarily conformal since it must have vanishing $\tau$.

A good candidate for a hydrodynamic-frame-invariant entropy current is
\begin{equation}
\label{entropy-LL}
\text{S}_{\mathbf{0}}=s_{\text{LL}} \text{u}_{\text{LL}} = 2\sqrt{\sigma \varepsilon_{\text{LL}}} \text{u}_{\text{LL}},
\end{equation}  
which can be expressed in any frame using Eqs. \eqref{eigene} and \eqref{eigenU}. This is usually adopted as the entropy current of a perfect fluid, and in that case it is divergence-free when external forces vanish. Here, it obeys (see \eqref{T-cons-el-mag-LL})
\begin{equation}
\nabla\cdot \text{S}_{\mathbf{0}}=-\sqrt{\frac{\sigma}{\varepsilon_\text{LL}}}\left(\Theta_\text{LL}\tau+\text{u}_\text{LL}\cdot \text{f}\right)=-\frac{1}{T_\text{LL}}\left(\Theta_\text{LL}\tau+\text{u}_\text{LL}\cdot \text{f}\right)\, ,
\end{equation}
which can be recast in terms of arbitrary-frame data using the already quoted  \eqref{eigene}, \eqref{eigenU} and the divergence of the latter. Expanding this result up to first order for $\chi, \tau\ll \varepsilon$, we find for a free fluid
\begin{equation}
\nabla\cdot \text{S}_{\mathbf{0}(1)}=-\frac{1}{T}\Theta \tau= \frac{\zeta}{T}\Theta^2
\, ,
\end{equation}
where we have used in the last equality  the first-order derivative expansion of $\tau$, given in \eqref{e1}. For this to be  positive one finds the usual requirement $\zeta>0$. From this perspective,
the current $\text{S}_{\mathbf{0}}$ seems fine. 

The expansion of $\text{S}_{\mathbf{0}}$ up to second order in $\chi, \tau\ll \varepsilon$,
\begin{equation}
\label{entropy-LL-disexp}
\text{S}_{\mathbf{0}}=2 \sqrt{\sigma\varepsilon}\text{u}+\chi \sqrt{\frac{\sigma}{\varepsilon}}\ast \text{u}-\frac{\chi^2}{4\varepsilon } \sqrt{\frac{\sigma}{\varepsilon}}\text{u}-\frac{\tau \chi}{2\varepsilon }\sqrt{\frac{\sigma}{\varepsilon}} \ast \text{u}+\cdots=
s\text{u}+\frac{\text{q}}{T} -\frac{\chi^2}{4\varepsilon T}\text{u}-\frac{\tau }{2\varepsilon T}\text{q}+\cdots,
\end{equation}  
is in agreement with the usual expectations dictated by \emph{extended irreversible thermodynamics} (completing the first-order \emph{classical irreversible thermodynamics})  \cite{RZ}. These can be summarized as follows, the order referring to the  dissipative expansion:
\begin{enumerate}
\item free perfect limit:
$\left.\text{S}\right\vert_{\chi=\tau=0}=\text{S}_{(0)}=s \text{u}=2 \sqrt{\sigma\varepsilon}\text{u}$;
\item stability
$\left.\frac{\partial\text{S}\cdot \text{u}}{\partial \tau}\right\vert_{\chi=\tau=0}=0 $;
\item first-order (CIT) correction: $\text{S}_{(1)}=\frac{\text{q}}{T} 
$;
\item second-order (EIT) corrections: $\text{S}_{(2)}$ might contain $\frac{\tau^2}{\varepsilon T} \text{u}$, $\frac{\chi^2}{\varepsilon T} \text{u}$ and $\frac{\tau}{\varepsilon T} \text{q}$;
\item second law: $\nabla \cdot \text{S}\geqslant 0$.
\end{enumerate}
Other invariant terms may be considered in the definition of $\text{S}$ as long as the above requirements are satisfied. In the absence of a concrete proposal for selecting other terms, we will not pursue the argument any further. Related discussions can be found in \cite{Bhattacharya:2011tra, Eling:2013bj,Banerjee:2013qha, Banerjee:2014ita}.\footnote{It should be quoted that $\text{S}$ as defined in \eqref{entropy-LL} does not coincide with the entropy current proposed in Ref. \cite{Banerjee:2014ita}. Hydrodynamic-frame invariance and CIT/EIT arguments were not part of the agenda in this work, based essentially on the second law of thermodynamics.} 

\subsubsection*{Light-cone versus Randers--Papapetrou frames}

\paragraph{Light-cone frame} Every two-dimensional metric is amenable by diffeomorphisms to a conformally flat form. This suggests to use:\footnote{With this choice, $g_{+-}=\nicefrac{1}{2}\,\text{e}^{-2\omega}$,  $\eta_{+-}=\nicefrac{1}{2}\,\text{e}^{-2\omega}$,
$\eta^{+-}=-2\text{e}^{2\omega}$, $\eta_{+}^{\hphantom{+}+}=1$, $\eta_{-}^{\hphantom{-}-}=-1$. Notice also that $\ast \left(\text{d}x^+\wedge\text{d}x^-\right)=\eta^{+-}=-2\text{e}^{2\omega}$.}
\begin{equation}
\label{carflc}
\text{d}s^2 =\text{e}^{-2\omega}\text{d}x^+\text{d}x^-
\end{equation}
(with usual time and space coordinates defined as $x^\pm=x\pm kt$), where $\omega$ is an arbitrary function of $x^+$ and $x^-$.

Any normalized congruence has the following form:
\begin{equation}
\label{flat-vel}
\text{u}= u_{+}\text{d}x^{+}+u_{-}\text{d}x^{-}\quad \Leftrightarrow \quad
\ast\text{u}=- u_{+}\text{d}x^{+}+u_{-}\text{d}x^{-},
\end{equation}
where $u_{\pm}$, functions of $x^+$ and $x^-$, are related by the normalization condition
\begin{equation}
\label{flat-vel-norm}
u_{+}u_{-}=-\frac{k^2}{4}\text{e}^{-2\omega}.
\end{equation}
We can parameterize the velocity field as
\begin{equation}
\label{flat-vel-comp}
u_{+}=-\frac{k}{2}\text{e}^{-\omega}\sqrt{\xi},\quad u_{-}=\frac{k}{2}\text{e}^{-\omega}\frac{1}{\sqrt{\xi}},
\end{equation}
where $\xi=\xi(x^+,x^-)$ is defined as the ratio 
\begin{equation}
\xi=-\frac{u_{+}}{u_{-}}.
\end{equation}
The choice  $\xi=1$ corresponds to a comoving fluid because in this case $\text{u}=-k^2\text{e}^{-\omega}\text{d}t$.

For the congruence at hand 
\begin{equation}
\Theta\pm \Theta^{\ast}=\pm 2k \text{e}^{2\omega}
\partial_\pm  \text{e}^{-\left(\omega\pm \ln \sqrt{\xi}\right)}.
\end{equation}
We can also determine the Weyl connection and field strength:
\begin{equation}
\label{Weylcccf}
\text{A}=  -\text{d}\omega +\ast\text{d} \ln \sqrt{\xi}
\quad \text{and}\quad
F=-\Box \ln \sqrt{\xi}=-2\text{e}^{2\omega}\partial_+\partial_- \ln \xi
,
\end{equation}
whereas the ordinary (non Weyl-covariant) scalar curvature reads (see \eqref{Wcurv})

\begin{equation}
\label{Rsclalarccf}
R=2\Box\omega
=8\text{e}^{2\omega}\partial_+\partial_- \omega
.
\end{equation}

In the present light-cone frame $\{\text{d}x^+,\text{d}x^-\}$,  a general energy--momentum tensor with $\epsilon=p$ has components
\begin{equation}
\begin{split}
T_{++}=
\frac{\xi}2\left(\varepsilon -\chi+ \frac{\tau}{2}\right){\text{e}^{-2\omega}}
,&\quad
T_{--}=\frac1{2\xi}\left(\varepsilon +\chi+ \frac{\tau}{2}\right)\text{e}^{-2\omega}
,\\
T_{+-}=T_{-+}&=\frac{\tau}{4}\text{e}^{-2\omega}.
 \end{split}
\end{equation}
For a conformal fluid Eqs. \eqref{confluidtau} lead to $T_{+-}=T_{-+}=0$ and 
\begin{equation}
\label{flat-vel-em}
(\varepsilon+\chi)(\varepsilon-\chi)=4\text{e}^{4\omega}T_{++}T_{--},
\quad
\frac{\varepsilon+\chi}{\varepsilon-\chi}=\frac{T_{--}}{T_{++}}\xi^2
.
\end{equation}
In the latter case, and in the absence of external forces, the forms \eqref{T-cons-el-mag-sol} are closed, which in light-cone coordinates implies
that $(\varepsilon-\chi){\text{e}^{-2\omega}}{\xi}$ is locally a function of $x^+$, and 
$(\varepsilon+\chi)\frac{\text{e}^{-2\omega}}{\xi}$ a function of $x^-$.
Observe that in the Landau--Lifshitz frame ($\chi_{\text{LL}}=0$)
\begin{equation}
\label{flat-vel-LL}
 \xi_{\text{LL}}^2 =\frac{T_{++}}{T_{--}}, \quad
 \varepsilon_{\text{LL}}^2 =4\text{e}^{4\omega}T_{++}T_{--}.
\end{equation}
In this frame, on-shell, $F$ vanishes. Moving from a given hydrodynamic frame to another by a local Lorentz boost, amounts to perform the following transformation on the function $\xi$
\begin{equation}
\label{hydroboostxi}
 \xi(x^+,x^-)\to\xi^\prime(x^+,x^-)= \text{e}^{-2\psi(x^+,x^-)} \xi(x^+,x^-).
\end{equation}

\paragraph{Randers--Papapetrou frame} The light-cone frame is not well suited for the Carrollian limit, which is the ultra-relativistic limit reached at vanishing $k$, and emerging at the null-infinity conformal boundary of a flat spacetime (subject of next section). As discussed in \cite{CMPPS1}, Carrollian fluid dynamics is elegantly reached in the Randers--Papapetrou frame, where 
\begin{equation}
\label{carrp}
\text{d}s^2 =- k^2\left(\Omega \text{d}t-b_{x} \text{d}x
\right)^2+a \text{d}x^2
\end{equation}
with all three functions of the coordinates $t$ and $x$. 

A generic velocity  vector field $\text{u}$ reads:
\begin{equation}
\label{ut}
\text{u}= \gamma\left(\partial_t +v^x \partial_x
\right).
\end{equation}
It is convenient to parametrize the velocity $v^x$ (see \cite{CMPPS1}) as\footnote{With these definitions, $\beta^x$ transforms as the component of a genuine Carrollian vector $\pmb{\beta}=\beta^x \partial_x$, when considering the flat limit of the bulk spacetime.
Notice that
$\beta_x+b_x=-\frac{\Omega u_x}{ku_0}$. We define as usual $b^x=a^{xx}b_x$, $\beta_x = a_{xx}\beta^x$, $v_x = a_{xx}v^x$ with $a_{xx}=\nicefrac{1}{a^{xx}}= a$, 
$\pmb{b}^2=b_xb^x$,  $\pmb{\beta}^2=\pmb{\beta}\cdot \pmb{\beta}=\beta_x \beta^x$ and $\pmb{b}\cdot \pmb{\beta}=b_x \beta^x$.}
\begin{equation}
\label{vbeta}
v^{x}=\frac{k^2\Omega\beta^x}{1+k^2\pmb{\beta}\cdot\pmb{b}}
\Leftrightarrow
\beta^{x}=\frac{v^x}{k^2\Omega\left(1-\frac{v^x b_x}{\Omega}\right)}
\end{equation}
with  Lorentz factor 
\begin{equation}
\gamma =\dfrac{1+k^2 \pmb{\beta}\cdot\pmb{b}}{\Omega\sqrt{1-k^2\pmb{\beta}^2}}.
\end{equation}
The velocity form and its Hodge-dual read:
\begin{equation}
\label{uf}
\text{u}=- \frac{k^2}{\sqrt{1-k^2\pmb{\beta}^2}}\left(\Omega\text{d}t
-\left(b_x+\beta_ x\right)\text{d}x\right), \quad \ast\text{u}=k  \sqrt{a} \Omega \gamma\left( \text{d}x-v^x\text{d}t\right),
\end{equation}
while the corresponding vector is 
\begin{equation}
 \ast\text{u}= \frac{k}{\sqrt{a}\sqrt{1-k^2\pmb{\beta}^2}}\left(\frac{b_x+\beta_x}{\Omega}\partial_t+\partial_x\right).
\end{equation}

We can determine the form of the heat current $\text{q}$, which must be proportional to  $\ast\text{u}$,
in terms of a single component $q_x$. We find
\begin{equation}
\label{qchi}
\chi=\frac{q_x}{k\sqrt{a}\Omega \gamma}=\frac{q^x\sqrt{a}\sqrt{1-k^2\pmb{\beta}^2}}{k}.
\end{equation}
Similarly, for the viscous stress tensor
\begin{equation}
\label{tautau}
\tau=\frac{\tau_{xx}}{a\Omega^2 \gamma^2}=\tau^{xx} a \left(1-k^2\pmb{\beta}^2\right).
\end{equation}

Performing a local Lorentz boost  \eqref{locLor} on the hydrodynamic frame does not affect the geometric objects $\Omega$, $b_x$ or $a$, and is thus entirely captured by the transformation of the vector $\pmb{\beta}$. Parameterizing the boost in terms of a Carrollian vector $\pmb{B}=B^x \partial_x$ as 
\begin{equation}
\cosh \psi=\Gamma=\frac{1}{\sqrt{1-k^2\pmb{B}^2}},
\quad 
\sinh \psi=\Gamma k \sqrt{a}B^x=\frac{k \sqrt{a}B^x}{\sqrt{1-k^2\pmb{B}^2}}, 
\end{equation}
we get:
\begin{equation}
\label{betacompo}
\pmb{\beta}^\prime
=\frac{\pmb{\beta}+\pmb{B}}{1+k^2\pmb{\beta}\cdot \pmb{B}}
,
\end{equation}
as expected from the velocity rule composition in special relativity. Using \eqref{locLor-en-he-nc}, we also obtain
\begin{eqnarray}
\label{vare-trans}
\varepsilon^\prime&=&\frac{1}{1-k^2\pmb{B}^2}\left(
\left(1+k^2\pmb{B}^2\right)
\varepsilon-
k \sqrt{a}B^x
2\chi+k^2\pmb{B}^2\tau\right),
\\
\label{chi-trans}
\chi^\prime&=&\frac{1}{1-k^2\pmb{B}^2}\left(
\left(1+k^2\pmb{B}^2\right)
\chi-
k \sqrt{a}B^x
(2\varepsilon+\tau)\right),
\end{eqnarray}
accompanying \eqref{locLor-tau}.
Together with \eqref{qchi} and \eqref{tautau}, we finally reach:
\begin{eqnarray}
\label{qchiprime}
\frac{q_x^\prime}{\sqrt{a}}&=&\left(
\left(1+k^2\pmb{B}^2\right)
\chi-
k \sqrt{a}B^x
(2\varepsilon+\tau)\right)k
\frac{\left(1+k^2\left(\pmb{\beta}\cdot
\pmb{B}+\left(\pmb{\beta}+
\pmb{B}\right)\cdot\pmb{b}
\right)
\right)}{\left(1-k^2\pmb{\beta}^2
\right)^{\nicefrac{1}{2}}\left(1-k^2\pmb{B}^2
\right)^{\nicefrac{3}{2}}},\\
\label{tautauprime}
\frac{\tau_{xx}^\prime}{a}&=&\tau 
\frac{\left(1+k^2\left(\pmb{\beta}\cdot
\pmb{B}+\left(\pmb{\beta}+
\pmb{B}\right)\cdot\pmb{b}
\right)
\right)^2}{\left(1-k^2\pmb{\beta}^2
\right) \left(1-k^2\pmb{B}^2
\right)}.
\end{eqnarray}

\subsection{Carrollian fluids} \label{sec:carfluid}

\subsubsection*{The  Carrollian geometry}

The Carrollian geometry $\mathbb{R}\times \mathscr{S}$ is obtained as the vanishing-$k$ limit of the two-dimensional pseudo-Riemannian geometry $ \mathscr{M}$ equipped with metric \eqref{carrp}. In this limit, the line 
$\mathscr{S}$ inherits a metric\footnote{This metric lowers all $x$ indices.}
\begin{equation}
\label{dmet}
\text{d}\ell^2=a \text{d}x^2,
\end{equation}
and $t\in \mathbb{R}$ is the Carrollian time. Much like a Galilean space is observed from a spatial frame moving with respect to a local inertial frame with velocity $\mathbf{w}$, a Carrollian frame is described by a form $\pmb{b}=b_{x}(t,x)\, \text{d}x$.  The latter is \emph{not} a velocity because in Carrollian spacetimes motion is forbidden. It is rather an inverse velocity, describing a ``temporal frame'' and plays a dual role. A scalar $\Omega(t,x )$ also remains in the $k\to 0$ limit 
(as in the Galilean case, see \cite{CMPPS1} -- this reference will be useful along the present section). 

We define the Carrollian diffeomorphisms as
\begin{equation}
\label{cardifs} 
t'=t'(t,x)\quad \text{and} \quad x^{\prime}=x^{\prime}(x).
\end{equation}
The ordinary exterior derivative of a scalar function does not transform as a form. To overcome this issue, it is desirable to introduce a Carrollian derivative as 
\begin{equation}
\label{dhat}
\hat\partial_x=\partial_x+\frac{b_{x}}{\Omega}\partial_t,
\end{equation}
transforming as a form. With this derivative we can proceed and define a Carrollian covariant derivative $\hat \nabla_x$, based on Levi--Civita--Carroll connection
\begin{equation}
\label{dgammaCar}
\hat\gamma^x_{xx}= \hat\partial_x \ln\sqrt{a}.
\end{equation}

As we will see in \ref{sec:reconflat}, in the framework of flat holography, the spatial surface  $\mathscr{S}$ emerges as the null infinity $\mathscr{I}^+$ of the Ricci-flat geometry. The geometry of $\mathscr{I}^+$ is equipped with a conformal class of metrics rather than with a metric. From a representative of this class, we must be able to explore others by Weyl transformations, and this amounts to study conformal Carrollian geometry as opposed to plain Carrollian geometry (see \cite{Duval:2014uoa}). 

The action of Weyl transformations on the elements of the Carrollian geometry on a surface $\mathscr{S}$ is inherited from \eqref{conmet}
 \begin{equation}
 \label{weyl-geometry}
a\to\frac{a}{{\cal B}^2}, \quad b_{x}\to \frac{b_{x}}{{\cal B}}, 
\quad 
\Omega \to \frac{\Omega}{{\cal B}}, \quad \beta_{x}\to \frac{\beta_{x}}{{\cal B}}, 
\end{equation}
where $\mathcal{B}=\mathcal{B}(t,x)$ is an arbitrary function.  However, the Levi--Civita--Carroll covariant derivatives are not covariant under \eqref{weyl-geometry}. 
Following \cite{CMPPS1}, they must be replaced 
with
Weyl--Carroll covariant spatial and time metric-compatible derivatives built on the Carrollian acceleration $\varphi_x$ 
and the Carrollian expansion $\theta$,
\begin{eqnarray}
\label{caracc}
&\varphi_x=\dfrac{1}{\Omega}\left(\partial_t b_{x}+\partial_x \Omega\right)
=\partial_t \dfrac{b_{x}}{\Omega}+\hat\partial_x \ln \Omega
, &\\
\label{carshexp-tempcon}
&\theta=
\dfrac{1}{\Omega}  \partial_t \ln\sqrt{a}, &
\end{eqnarray}
which transform as connections:
 \begin{equation}
 \label{weyl-geometry-2-abs}
\varphi_{x}\to \varphi_{x}-\hat\partial_x\ln \mathcal{B},\quad \theta\to \mathcal{B}\theta-\frac{1}{\Omega}\partial_t \mathcal{B}.
\end{equation} 
In particular, these can be combined in\footnote{Contrary to $\varphi_{x}$, $\alpha_x$ is not a Carrollian one-form, \emph{i.e.} it does not transform covariantly under Carrollian diffeomorphisms \eqref{cardifs}.}
\begin{equation}
 \label{weyl-geometry-alpha}
\alpha_x=\varphi_{x}-\theta b_{x},
\end{equation} 
transforming under Weyl rescaling as 
\begin{equation} 
\label{weyl-geometry-alpha-trans}
\alpha_x\to \alpha_{x}-\partial_x\ln \mathcal{B}.
\end{equation} 

The spatial Weyl--Carrol derivative is
\begin{equation}
\label{CWs-Phi}
\hat{\mathscr{D}}_x \Phi=\hat\partial_x \Phi +w \varphi_x \Phi,
\end{equation}
for a weight-$w$ scalar function  $\Phi$, and 
\begin{equation}
\label{CWV}
\hat{\mathscr{D}}_x V^x=\hat\nabla_x V^x +(w-1) \varphi_x V^x ,
\end{equation}
for a vector with weight-$w$ component $V^x$. It does not alter the conformal weight, and is generalized to any tensor by Leibniz rule.  

Similarly we define the temporal Weyl--Carroll derivative by its action on a weight-$w$ function $\Phi$
\begin{equation}
\label{CWtimecovdersc}
\frac{1}{\Omega}\hat{\mathscr{D}}_t \Phi=
\frac{1}{\Omega}\partial_t \Phi +w \theta \Phi,
\end{equation}
which is a scalar of weight $w+1$ under \eqref{weyl-geometry}. Accordingly, the action of the Weyl--Carroll time derivative on a weight-$w$ vector is 
\begin{equation}
\label{CWtimecovdervecform}
\frac{1}{\Omega}\hat{\mathscr{D}}_t V^x=
\frac{1}{\Omega}\partial_t V^x +w\theta V^x
 .
\end{equation}
This is the component of a genuine Carrollian vector of weight $w+1$, and Leibinz rule allows  to generalize this action to any tensor. 

The Weyl--Carroll connections have curvature. Here, the only non-vanishing piece is the curvature one-form resulting from the commutation of $\hat{\mathscr{D}}_x$ and $\frac{1}{\Omega}\hat{\mathscr{D}}_t$, which has weight $1$:
\begin{equation}
\mathscr{R}_x=\frac{1}{\Omega}\left(
\partial_t\alpha_x
-
\partial_x(\theta\Omega)
\right)=
\frac{1}{\Omega}\partial_t\varphi_x-\theta\varphi_x
-\hat\partial_x\theta
.\label{carcurv}
\end{equation}

 \subsubsection*{Carrollian fluid observables}

A relativistic fluid satisfying Eq. \eqref{T-cons} will obey Carrollian dynamics in the ultra-relativistic limit, reached at vanishing $k$. 
The original relativistic fluid is not at rest, but has a velocity parametrized with $\pmb{\beta}=\beta_x\text{d}x$ (see \eqref{vbeta}), which remains in the Carrollian limit as
 the kinematical ``inverse-velocity'' variable. We will keep calling it abusively ``velocity''. This variable transforms as a Carrollian vector and allows to define further kinematical objects. 
\begin{itemize}
\item We introduce the acceleration $\pmb{\gamma}=\gamma_x \text{d}x$
 \begin{equation}
 \label{car-gam}
\gamma_x=\dfrac{1}{\Omega} \partial_t \beta_x.
\end{equation}
This is not Weyl-covariant, as opposed to
 \begin{equation}
 \label{car-del}
\delta_x=\frac{1}{\Omega}\hat{\mathscr{D}}_t\beta_x=\gamma_x- \theta \beta_x= \dfrac{\sqrt{a}}{\Omega} \partial_t \frac{\beta_x}{\sqrt{a}},
\end{equation}
which has weight $0$.
\item The suracceleration is the weight-$1$ conformal Carrollian one-form
\begin{equation}
\mathscr{A}_x=\frac{1}{\Omega}\hat{\mathscr{D}}_t\frac{1}{\Omega}\hat{\mathscr{D}}_t\beta_x=\frac{1}{\Omega}\partial_t\left(\frac{1}{\Omega}\partial_t\beta_x-\theta\beta_x\right)
.
\label{carcsurac}
\end{equation}
It can be combined with the curvature \eqref{carcurv}, which has equal weight,
\begin{equation}
\label{car-surac}
s_x=\mathscr{A}_x+\mathscr{R}_x=\frac{1}{\Omega}\partial_t\left(\frac{1}{\Omega}\partial_t\beta_x-\theta\beta_x\right)+\frac{1}{\Omega}\partial_t\varphi_x-\theta\varphi_x
-\hat\partial_x\theta.
\end{equation} 
This appears as a conformal Carrollian total (\emph{i.e.}  kinematical
plus geometric) suracceleration, and enables us to define a weight-$2$ conformal Carrollian scalar:
\begin{equation}
\label{dyn-WCc}
s=\frac{s_x}{\sqrt{a}}.
\end{equation} 
The latter originates from the Weyl curvature $F$ of the pseudo-Riemannian ascendent manifold~$\mathscr{M}$:
 \begin{equation}
s=-\lim_{k \to 0} kF.
\end{equation} 
Notice that the ordinary scalar curvature of $\mathscr{M}$ given in \eqref{Wcurv} is not Weyl-covariant (see \eqref{anomR}) and can be expressed in terms of Carrollian non-Weyl-covariant scalars of $\mathbb{R}\times \mathscr{S}$:
 \begin{equation}
\label{R-carform}
R=\frac{2}{k^2}\left(\theta^2+\frac{1}{\Omega}\partial_t\theta\right)-2\left(\hat\nabla_x +\varphi_x\right)\varphi^x.
\end{equation} 

\end{itemize}

Besides the inverse velocity, acceleration and suracceleration, other physical data describe a Carrollian fluid.
\begin{itemize}
\item The energy density $\varepsilon$ and the pressure $p$, related here through  $\varepsilon=p$. The Carrollian energy and pressure are the zero-$k$ limits of the corresponding relativistic quantities, and have weight $2$. It is implicit that they are finite, and
in order to avoid inflation of symbols, we have kept the same notation.
\item The heat current
$\pmb{\pi}=\pi_{x} (t,x)  \text{d}x$ of conformal weight~$1$, inherited from the relativistic heat current (see \eqref{T}) as follows:\footnote{In arbitrary dimensions one generally admits  $q^{x}=Q^{x}+k^2 \pi^{x} +\text{O}\left(k^4\right)$ (see \cite{CMPPS1}), which amounts assuming
 $\chi=\frac{\chi_Q}{k}+\chi_\pi k + \text{O}\left(k^3\right)$. This is actually more natural because vanishing $\chi_Q$ is not a hydrodynamic-frame-invariant feature in the presence of friction.  
Keeping  $\chi_Q\neq 0$, however, is not viable holographically in two boundary dimensions because it would create a $\nicefrac{1}{k^2}$ divergence inside the derivative expansion. Since the Carrollian limit destroys anyway the hydrodynamic-frame invariance, our choice is consistent from every respect. 
Ultimately these behaviours  should be justified  within a microscopic quantum/statistical approach, missing at present.}
\begin{equation}
\label{QexpC}
q^{x}=k^2 \pi^{x} +\text{O}\left(k^4\right).
\end{equation}
This translates the expected (see \eqref{qchi}) small-$k$  behaviour  of  $\chi$:
\begin{equation}
\label{chidec}
\chi=\chi_\pi k + \text{O}\left(k^3\right),
\end{equation}
leading to
\begin{equation}
\label{Qexpchi}
\pi^x=\frac{\chi_\pi }{\sqrt{a}}.
\end{equation}
\item The 
weight-$0$ viscous stress tensors
 $\pmb{\Sigma}=\Sigma_{xx} \text{d}x^2$
 and
  $\pmb{\Xi}=\Xi_{xx} \text{d}x^2$, obtained from the relativistic viscous stress tensor $
 \frac{\tau}{k^2} \ast \text{u}\ast \text{u}$
as
\begin{equation}
\label{XiexpC}
\tau^{xx}=-\frac{\Sigma^{xx}}{k^2}-\Xi^{xx}+\text{O}\left(k^2\right).
\end{equation}
For this to hold, following \eqref{tautau}, we expect 
\begin{equation}
\label{taudec}
\tau=\frac{\tau_\Sigma}{k^2}+ \tau_\Xi+\text{O}\left(k^2\right),
\end{equation}
and find (in the Carrollian geometry, indices are lowered with $a_{xx}=a$):
\begin{equation}
\label{Xiexpchi}
\Sigma^{x}_{\hphantom{x}x}=-\tau_\Sigma,\quad
\Xi^{x}_{\hphantom{x}x}=-\tau_\Xi-\pmb{\beta}^2 \tau_\Sigma.
\end{equation}
As we will see later, this is in agreement with the form of $\tau$ for the relativistic systems at hand (see Eqs. \eqref{R-carform} and \eqref{anomaly}).
\item Finally, we assume that the components of the external force density behave as follows, providing further Carrollian power and tension:
\begin{equation}
\label{f}
\begin{cases}
\frac{k}{\Omega}f_0=\frac{f}{k^2}+e+\text{O}\left(k^2\right),
\\
f^x=
\frac{h^x}{k^2}+g^x
+\text{O}\left(k^2\right).
\end{cases}
\end{equation}
\end{itemize}

 \subsubsection*{Hydrodynamic equations}

The hydrodynamic equations for a Carrollian fluid are obtained as the zero-$k$ limit of the relativistic equations (see \cite{CMPPS1}):
\begin{eqnarray}
-\left(\frac{1}{\Omega}\partial_t +2\theta\right)\left(\varepsilon
-\pmb{\beta}^2\Sigma^{x}_{\hphantom{x}x}
\right)
+\left(\hat\nabla^x +2\varphi^x \right)
\left(\beta_x
\Sigma^{x}_{\hphantom{x}x}
\right)
+
\theta\left(
\Xi^{x}_{\hphantom{x}x}-\pmb{\beta}^2\Sigma^{x}_{\hphantom{x}x}\right)
&=& e
,
 \label{carE} 
\\
\theta \Sigma^{x}_{\hphantom{x}x}
&=& f
,
 \label{carF} 
\\
\left(\hat\nabla_x+\varphi_x\right)\left(\varepsilon-\Xi^{x}_{\hphantom{x}x}
\right) +\varphi_x \left(\varepsilon-\pmb{\beta}^2\Sigma^{x}_{\hphantom{x}x} \right)+\left(\dfrac{1}{\Omega}\partial_t+\theta \right)
\left(\pi_x+\beta_x\left(2\varepsilon -\Xi^{x}_{\hphantom{x}x}\right)
\right) &=&g_x,\qquad
  \label{carG} 
 \\
 -\left(\hat\nabla_x+\varphi_x\right)\Sigma^{x}_{\hphantom{x}x} 
-\left(\frac{1}{\Omega}\partial_t +\theta
\right)\left(\beta_x
\Sigma^{x}_{\hphantom{x}x}
\right)
 &=&h_x
.
 \label{carH} 
\end{eqnarray}

Generically, the above equations are not invariant under Carrollian local boosts, acting as 
\begin{equation}
\label{betacompocar}
\beta_x^\prime
=\beta_x+B_x
\end{equation} 
(vanishing-$k$ limit of \eqref{betacompo}). This should not come as a surprise. Such an invariance is exclusive to the relativistic case for obvious physical reasons, and is also known to be absent from Galilean fluid equations, which are not invariant under local Galilean boosts.  Nevertheless, as we will see in Sec. \ref{sec:recon-flat-bry}, in specific situations a residual invariance persists.

\section{Three-dimensional bulk reconstruction}\label{sec:recon}

\subsection{Anti-de Sitter}\label{sec:reconAdS}

Three-dimensional Einstein spacetimes are peculiar because the usual derivative expansion terminates at finite order. This happens also for the Fefferman--Graham expansion (see \emph{e.g.} \cite{skesol}). The reason is that most geometric and fluid tensors vanish (like the shear or the vorticity), reducing the number of available terms compatible with conformal invariance.
Indeed, following the original fluid/gravity works \cite{Bhattacharyya:2007, Haack:2008cp, Bhattacharyya:2008jc, Hubeny:2011hd}, the ansatz for the bulk Einstein metric is a power expansion in $\nicefrac{1}{r}$ such that boundary Weyl transformations \eqref{conmet} are compensated by $r\to \mathcal{B}(t,x) r$.
The boundary metric has weight $-2$, the forms $\text{u}$ and $\ast \text{u}$ (velocity and dual fluid velocity) weight $-1$, whereas the energy and heat densities of the fluid have weight $2$. The Weyl connection $\text{A}$ has (anomalous) weight zero, as the form $\text{d}r$.  With these data
we obtain:
\begin{equation}
\boxed{
\text{d}s^2_{\text{Einstein}} =
2\frac{\text{u}}{k^2}\left(\text{d}r+r \text{A}\right)+r^2\text{d}s^2+ \frac{8\pi G}{k^4} \text{u}\left(\varepsilon \text{u}  +\chi \ast \text{u} \right),}
\label{papaefgenrescrec3d}
\end{equation}
where $\text{A}$ is displayed in \eqref{Wconc}, $\varepsilon$ and $\chi$ being 
the energy and heat densities of the fluid (as opposed to higher dimension, the heat current appears explicitly in the ansatz). 
These enter the fluid energy--momentum tensor \eqref{Tgen} together with $\tau$, which carries the anomaly: 
\begin{equation}
\label{anomaly}
\tau=\frac{R}{8\pi G}=
\frac{1}{4\pi Gk^2}\left(
\Theta^2- \Theta^{\ast 2}
+\text{u}(\Theta) - \ast \text{u}(\Theta^\ast) 
\right)
\end{equation}
(we keep the conformal state equation $\varepsilon = p$).
For a flat boundary this trace is absent, but Weyl transformations bring it back.

The precise coefficients of the eligible terms in the ansatz are determined by the radial-evolution subset of Einstein's equations, and this is already taken care of in expression \eqref{papaefgenrescrec3d}, utterly locking the $r$-dependence. The remaining Einstein's equations further constrain the boundary data, \emph{i.e.} the metric and the fluid. Summarizing, the metric \eqref{papaefgenrescrec3d} provides an \emph{exact} Einstein, asymptotically AdS spacetime, 
with $R=6\Lambda = -6k^2$, under the necessary and sufficient condition that the non-conformal fluid energy--momentum tensor \eqref{Tgen} obeys
\begin{equation}
\label{T-tilde-cons}
\nabla^\mu \left(T_{\mu\nu}+D_{\mu\nu}\right)=0,
    \end{equation}
where $D_{\mu\nu}$ is a symmetric and traceless tensor which reads:
\begin{equation}
\label{D}
D_{\mu\nu}\text{d}x^\mu \text{d}x^\nu =\frac{1}{8\pi G k^4}
\left(\left(\text{u}(\Theta) + \ast \text{u}(\Theta^\ast)-\frac{k^2}{2}R\right) 
\left(\text{u}^2+
 \ast \text{u}^2
\right)-4\ast\text{u}(\Theta)\text{u} \ast \text{u}
\right).
\end{equation}
On the one hand, the holographic energy--momentum tensor is the sum  $T_{\mu\nu}+D_{\mu\nu}$, and this can be shown following the Balasubramanian--Kraus method \cite{bala}.\footnote{For this computation we used the conventions of \cite{Caldarelli:2011wa}.}  On the other hand, the holographic fluid is subject to an external force with density
\begin{equation}
f_\nu=-\nabla^\mu D_{\mu\nu}.
\end{equation}
Its longitudinal and transverse components are
\begin{equation}
\label{force}
\begin{cases}
u^\mu f_\mu= -\frac{1}{4\pi G}
\left(\ast \text{u}(F)+2\Theta^\ast F +\frac{1}{2}
\Theta R\right),
\\
\ast u^\mu f_\mu= \frac{1}{8\pi G}\left(\ast \text{u}(R)+\Theta^\ast  R\right).
\end{cases}
\end{equation}
Combining \eqref{T-cons-el-mag-nc}, \eqref{anomaly} and \eqref{force} we find the following equations: 
\begin{equation}
\label{T-cons-el-mag-nc-force} 
 \begin{cases}
\left(u^\mu+\ast u^\mu  \right)\mathscr{D}_\mu \left(\varepsilon+
 \chi
\right)
=\frac{1}{4\pi G}\ast u^\mu\mathscr{D}_\mu F,
\\
\left(u^\mu-\ast u^\mu\right)\mathscr{D}_\mu \left(\varepsilon-
 \chi
\right)
=\frac{1}{4\pi G}\ast u^\mu\mathscr{D}_\mu F.
\end{cases}
\end{equation} 
Notice that eventually these equations are Weyl-covariant (weight-$3$) despite the conformal anomaly.

An important remark is in order regarding the holographic fluid. Rather than $T_{\mu\nu}$, we could have adopted $T_{\mu\nu}+D_{\mu\nu}$ as its energy--momentum tensor. The latter would have been decomposed as in \eqref{T}, with $\tilde \varepsilon=\tilde p$ and $\tilde \chi$ though ($\tilde \tau=\tau$ since $D_{\mu\nu}$ has vanishing trace):
\begin{eqnarray}
\label{tilvarep}
\tilde\varepsilon&=&\varepsilon
+\frac{1}{8\pi Gk^2}\left(\text{u}(\Theta) + \ast \text{u}(\Theta^\ast)\right) 
- \frac{R}{16\pi G},
\\
\label{tilvarchi}
\tilde\chi&=& \chi -\frac{1}{4\pi Gk^2}\ast\text{u}(\Theta)
.
\end{eqnarray}
We did not make this choice for two reasons: (\romannumeral1) in the formula \eqref{papaefgenrescrec3d} we used $\varepsilon$ and $\chi$ rather than  $\tilde\varepsilon$ and $\tilde\chi$ for reconstructing the bulk;  (\romannumeral2)  $\varepsilon$ and $\nicefrac{\chi}{k}$ are finite in the limit of vanishing $k$, whereas $\tilde\varepsilon$ and $\nicefrac{\tilde\chi}{k}$ are not. This last fact is not an obstruction, but it would require to reconsider the Carrollian hydrodynamic equations developed in Ref. \cite{CMPPS1} and applied here. 

Expression \eqref{papaefgenrescrec3d} is the most general locally AdS spacetime in Eddington--Finkelstein coordinates. The corresponding gauge includes but does not always coincide with BMS.\footnote{There is no definition of Eddington--Finkelstein gauge. Within the three-dimensional derivative expansion, one can nevertheless refer to it as a gauge because the $r$-dependence is fixed. This does not exhaust all freedom, but allows comparison with BMS. Actually, fluid/gravity approach is not meant to lock completely the coordinates for describing the most general solution in terms of a minimal set of functions. \label{EF-BMS}} From that perspective, this result is new although it may not contain any new solutions compared \emph{e.g.} to Ba\~nados' \cite{Banados_solutions}, all  
captured either in BMS or in Fefferman--Graham gauge (see \cite{Barnich:2010eb}).
The bonus is the hydrodynamical interpretation. Here the corresponding fluid is defined on a generally curved boundary and has an arbitrary velocity field. This should be contrasted with the treatment of three-dimensional fluid/gravity correspondence worked out in Refs. \cite{ Haack:2008cp, Bhattacharyya:2008jc}, where the host geometry was flat, avoiding the issue of conformal anomaly. Furthermore the fluid was assumed perfect by hydrodynamic-frame choice, which permits a subclass of Ba\~nados solutions only,  as we will see in Sec. \ref{sec:recon-flat-bry} by computing the conserved charges.

For practical purposes, we can work in light-cone coordinates, introduced in Eq. \eqref{carflc}. Using the expression \eqref{flat-vel-comp} for the congruence $\text{u}$, and solving the fluid equations \eqref{T-cons-el-mag-nc-force}, we obtain the fluid densities $\varepsilon$ and $\chi$ in terms of two arbitrary chiral functions $\ell_\pm(x^\pm)$:
\begin{eqnarray}
\label{gen-varepsilon}
\varepsilon 
&=&\frac{\text{e}^{2\omega}}{4\pi G}\left(\frac{\ell_+}{\xi}+\xi \ell_-
-\frac{3\left(\partial_+\xi\right)^2}{4\xi^3}
+\frac{\partial^2_+\xi}{2\xi^2}
+\frac{\left(\partial_-\xi\right)^2}{4\xi}
-\frac{\partial^2_-\xi}{2}\right),
\\
\label{gen-chi}
\chi &=&
\frac{\text{e}^{2\omega}}{4\pi G}\left(-\frac{\ell_+}{\xi}+\xi \ell_-+\frac{3\left(\partial_+\xi\right)^2}{4\xi^3}
-\frac{\partial^2_+\xi}{2\xi^2}
+\frac{\left(\partial_-\xi\right)^2}{4\xi}
-\frac{\partial^2_-\xi}{2}
+\frac{\partial_+\xi\partial_-\xi}{\xi^2}-\frac{\partial_+\partial_-\xi}{\xi}\right)
.\qquad
\end{eqnarray}
Gathering these data together with \eqref{Weylcccf} inside \eqref{papaefgenrescrec3d} 
provides, in the gauge at hand, the general class of locally AdS three-dimensional spacetime with curved conformal boundary. The conformal factor $\exp{2\omega}$ can be apparently reabsorbed 
by setting $r$ to $r \exp{\omega}$, thus bringing \eqref{papaefgenrescrec3d} to its flat-boundary form.\footnote{This should be contrasted with the more intricate situation regarding this conformal factor inside the analogous formula in Fefferman--Graham gauge,  Eq. (2.21) of Ref. \cite{Barnich:2010eb}.} 
One should nevertheless be careful when making claims based on coordinate redefinitions, even in seemingly safe situations, because they can potentially alter global properties. Indeed, as discussed in Ref. \cite{Troessaert:2013fma}, $\omega$ is expected to bring different asymptotics and new charges, and the corresponding solutions might generalize {\it Ba\~nados}' family. In our subsequent analysis of Sec. \ref{sec:recon-flat-bry-AdS}, we will set $\omega=0$.
As we will shortly see, the arbitrary function $\xi(x^+, x^-)$ is also insidious regarding the charges, and focusing on it will be sufficient for the scope of this work. 

We could proceed and display similar expressions in the Randers--Papapetrou boundary frame, describing the general locally anti-de Sitter spacetimes in terms of the three geometric data $\Omega(t,x)$, $b_x(t,x)$ and $a_{xx}=a(t,x)$, and whatever integration functions would appear in the process of solving the hydrodynamic equations \eqref{T-cons-el-mag-nc-force}. Usually, this resolution cannot be conducted explicitly as it happens in light-cone coordinates, and we end up with an implicit description of the bulk metric. We should quote here that a specific example of curved  boundary\footnote{In that case $\Omega=\exp 2\beta$, $b_x=0$, $a=1$ and, in our language, the fluid velocity would have been $\text{u}=-k^2\text{e}^{2\beta}\text{d}t$, \emph{i.e.} comoving.}  
was investigated in Ref. \cite{Fareghbal:2015bxd}, outside of the fluid/gravity framework, and the output agrees with our general results. We should also stress, following the discussion of footnote \ref{EF-BMS}, that the Randers--Papapetrou boundary frame produces in \eqref{papaefgenrescrec3d}  
order-$r$ $\text{d}t  \text{d}x$ components absent in the BMS gauge.

\subsection{Ricci-flat}\label{sec:reconflat}

Our starting point is the finite derivative expansion of an asymptotically $\text{AdS}_3$ spacetime, Eq. \eqref{papaefgenrescrec3d}.  The fundamental question is whether the latter admits a smooth zero-$k$ limit. 

We have implicitly assumed that the Randers--Papapetrou data of the two-dimensional pseudo-Riemannian conformal boundary $\mathscr{I}$ associated with the original Einstein spacetime, $a$, $b$ and $\Omega$, remain unaltered at vanishing $k$, providing therefore directly the Carrollian data for the new spatial one-dimensional boundary $\mathscr{S}$ emerging at $\mathscr{I}^+$.  Following again the detailed analysis performed in \cite{CMPPS1}, we can match the various two-dimensional Riemannian quantities with the corresponding one-dimensional Carrollian ones:
\begin{equation}
\text{u}= -k^2\left(\Omega\text{d}t-\left(b_x+\beta_x\right)\text{d}x\right) + \text{O}\left(k^4\right), \quad \ast\text{u}=k  \sqrt{a} \text{d}x + \text{O}\left(k^3\right)
\end{equation}
and
\begin{equation}
\begin{array}{rcl}
\Theta&=&\theta+ \text{O}\left(k^2\right),\\
\text{a}&=&k^2\left(\varphi_x + \gamma_x\right)\text{d}x+ \text{O}\left(k^4\right),\\
\text{A}&=&\theta\Omega \text{d}t + \left(\alpha_x+\delta_x\right)\text{d}x + \text{O}\left(k^2\right),
\end{array}
\label{riemcar}
\end{equation}
where the left-hand-side quantities are Riemannian, and the right-hand-side ones Carrollian
(see
\eqref{caracc},
\eqref{carshexp-tempcon},  \eqref{weyl-geometry-alpha}, \eqref{car-gam},  \eqref{car-del}).

The closed form 
\eqref{papaefgenrescrec3d}
is smooth at zero $k$. In this limit the metric reads:
\begin{equation}
\boxed{
\begin{split}
\text{d}s^2_{\text{flat}} =&
-2\left(\Omega\text{d}t-\pmb{b}-\pmb{\beta}\right)\left(\text{d}r+r \left(\pmb{\varphi}+\pmb{\gamma}+\theta \left( \Omega\text{d}t-\pmb{b}-\pmb{\beta}\right)\right)\right)\\
&
+r^2\text{d}\ell^2
+8\pi G \left(\Omega\text{d}t-\pmb{b}-\pmb{\beta}\right) \left(\varepsilon\left(\Omega\text{d}t-\pmb{b}-\pmb{\beta}\right)-\pmb{\pi}\right),
\end{split}}
\label{papaefresricf}
\end{equation}
Here
$\text{d}\ell^2$,  $\Omega$, $\pmb{b}=b_{x}\text{d}x$, $\pmb{\varphi}=\varphi_x\text{d}x$  and $\theta$ are the Carrollian geometric objects introduced earlier. The bulk Ricci-flat spacetime is now dual to a Carrollian fluid with  kinematics captured in $\pmb{\beta}=\beta_x\text{d}x$ and $\pmb{\gamma}=\gamma_x\text{d}x$, energy density $\varepsilon$  (zero-$k$ limit of the corresponding relativistic function), and heat current $\pmb{\pi}=\pi_{x}\text{d}x$ (obtained in Eqs.\eqref{QexpC}, \eqref{chidec} and  \eqref{Qexpchi}). 

For the fluid under consideration, there is also a pair of Carrollian stress tensors originating from the anomaly  \eqref{anomaly}.  Using expressions \eqref{R-carform} and \eqref{taudec}, we can determine $\tau_\Sigma$ and $\tau_\Xi$, and Eqs. \eqref{Xiexpchi} provide in turn the Carrollian stress: 
\begin{equation}
\label{Xiexpchi-anomaly}
\Sigma^{x}_{\hphantom{x}x}=-\frac{1}{4\pi G}\left(\theta^2+\frac{\partial_t\theta}{\Omega}\right)
,\quad
\Xi^{x}_{\hphantom{x}x}=\frac{1}{4\pi G}\left(
\left(\hat\nabla_x +\varphi_x\right)\varphi^x
-\pmb{\beta}^2
\left(\theta^2+\frac{\partial_t\theta}{\Omega}\right)
\right).
\end{equation}
This is the advertised Carrollian emanation of the relativistic conformal anomaly. 

Expression \eqref{papaefresricf} will grant by construction an exact Ricci-flat spacetime provided the conditions under which \eqref{papaefgenrescrec3d} was Einstein are fulfilled in the zero-$k$ limit. These are the set of Carrollian hydrodynamic equations \eqref{carE}, \eqref{carF},  \eqref{carG} and \eqref{carH}, with Carrollian power and force densities $e$, $f$, $g_x$, $h_x$ obtained using their definition \eqref{f} and the expressions of $f_\mu$ displayed in \eqref{force} (we use for this computation the expression of the scalar curvature \eqref{R-carform}, and $s_x$ as given in \eqref{car-surac}). Equations \eqref{carF} and \eqref{carH} are automatically satisfied, whereas  \eqref{carE} and \eqref{carG} lead to\footnote{We remind that  Weyl--Carroll covariant derivatives are defined in Eqs. \eqref{CWs-Phi}, \eqref{CWV}, \eqref{CWtimecovdersc} and \eqref{CWtimecovdervecform}. Here $\varepsilon$,  $\beta^x$, $\pi_x$ and $s^x$ have weights $2$, $1$, $1$ and $3$. For example $\hat{\mathscr{D}}_x s^{x}= \hat{\nabla}_x s^x+2 \varphi_x s^x=\frac{1}{\sqrt{a}}
\hat\partial_x (\sqrt{a} s^x)
+2\varphi_x s^x
$.}
\begin{equation}
 \label{flat3dfinbetasimple} 
\begin{cases}
\dfrac{1}{\Omega}\hat{\mathscr{D}}_t\varepsilon+\dfrac{1}{4\pi G}\left(\dfrac{2s_x}{\Omega}\hat{\mathscr{D}}_t\beta^x+  
 \dfrac{\beta_x}{\Omega}\hat{\mathscr{D}}_t s^x+\hat{\mathscr{D}}^x s_x\right)=0,
\\
\hat{\mathscr{D}}_x \varepsilon -\dfrac{\beta_x}{\Omega}\hat{\mathscr{D}}_t \varepsilon
+ \dfrac{1}{\Omega}\hat{\mathscr{D}}_t \left(\pi_{x}+2\varepsilon \beta_x \right)= 
0.
\end{cases}
\end{equation}
The unknown functions, which bear the fluid configuration, are $\varepsilon(t,x)$,  $\pi_x(t,x)$ and $\beta_x(t,x)$. These cannot be all determined by the two equations at hand. Hence, there is some redundancy, originating from the relativistic fluid frame invariance -- responsible \emph{e.g.} for the arbitrariness of $\xi(x^+,x^-)$ in the description of AdS spacetimes using the light-cone boundary frame. More will be said about this in Sec. \ref{sec:recon-flat-bry-flat}.

Equations  \eqref{flat3dfinbetasimple} 
are Carroll--Weyl covariant. The Ricci-flat line element \eqref{papaefresricf} inherits Weyl invariance from its relativistic ancestor. The set of transformations \eqref{weyl-geometry}, \eqref{weyl-geometry-2-abs} and \eqref{weyl-geometry-alpha-trans}, supplemented with  $\varepsilon\to \mathcal{B}^2 \varepsilon$ and $\pi_{x}\to \mathcal{B} \pi_{x}$, can indeed be absorbed by setting $r\to \mathcal{B}r$, resulting thus in the invariance of \eqref{papaefresricf}. In the relativistic case this invariance was due to the AdS conformal boundary. In the case at hand, this is rooted to the location of the one-dimensional spatial boundary $\mathscr{S}$ at null infinity $\mathscr{I}^+$.

We would like to close this chapter with a specific but general enough situation to encompass all 
Barnich--Troessaert Ricci-flat three-dimensional spacetimes.  The Carrollian geometric data are $b_x=0$, $\Omega=1$ and $a=\exp 2\Phi(t,x)$, and the kinematic variable of the Carrollian dual fluid $\beta_x$ is left free. 
Hence \eqref{papaefresricf} reads:
\begin{eqnarray}
\text{d}s^2_{\text{flat}}&=& - 2 \left(\text{d}t -\beta_x \text{d}x
\right)\left(\text{d}r+r\left(
\partial_t\Phi \text{d}t
+\left(\partial_t-\partial_t\Phi 
\right)\beta_x\text{d}x
\right)
\right) \nonumber
\\
&&+ r^2 \text{e}^{2\Phi} \text{d}x^2 + 8\pi G \left(\text{d}t -\beta_x \text{d}x
\right) \left( \varepsilon \text{d}t - \left(\pi_x+\varepsilon\beta_x\right) \text{d}x\right),
\label{glen-flat}
\end{eqnarray}
where $\varepsilon(t,x)$ and $\pi_x(t,x)$ obey Eqs. \eqref{flat3dfinbetasimple} in the form
\begin{equation}
 \label{flat3dfinbetasimple-glenn} 
\begin{cases}
\left(\pa_t+2\pa_t \Phi\right)\varepsilon+\dfrac{1}{4\pi G}\left(2s_x\left(\pa_t+\pa_t \Phi\right)\beta^x+  
 \beta_x\left(\pa_t+3\pa_t \Phi\right) s^x+\left(\pa_x+\pa_x\Phi\right) s^x\right)=0,
\\
\partial_x \varepsilon
+\left(\partial_t+ \partial_t \Phi 
\right)\pi_{x}+2\varepsilon \pa_t \beta_x+\beta_x \pa_t \varepsilon 
= 
0.
\end{cases}
\end{equation}
Here, $s_x$ takes the simple form
\beq
s_x=\pa^2_t \beta_x-\pa_t \left(\beta_x\pa_t\Phi\right)-\pa_t\pa_x\Phi.
\eeq
For vanishing $\beta_x$, the results \eqref{glen-flat} and \eqref{flat3dfinbetasimple-glenn} coincide precisely with those obtained in \cite{Barnich:2010eb} by demanding Ricci-flatness in the BMS gauge. 
Here, they are reached from purely Carrollian-fluid considerations, and for generic $\beta_x(t,x)$, the metric \eqref{glen-flat} lays outside the BMS gauge.

\section{Two-dimensional flat boundary and conserved charges}\label{sec:recon-flat-bry}

We will now restrict the previous analysis to Ricci-flat and Weyl-flat boundaries, both in AdS and Ricci-flat spacetimes. This enables us to compute the conserved charges following \cite{BarnichCG1, BarnichCG2, code}, and analyze the role of the velocity and the heat current of the boundary fluid. 

\subsection{Charges in AdS spacetimes}\label{sec:recon-flat-bry-AdS}

The flatness requirements are equivalent to setting $R=0$ and $ F=0$.
In the light-cone frame \eqref{carflc}, this amounts to (see \eqref{Weylcccf} and \eqref{Rsclalarccf})
\begin{equation}
\label{xixipxim}
\omega=0\quad \text{and}\quad \xi(x^+,x^-)=- \frac{\xi^-(x^-)}{\xi^+(x^+)},
\end{equation}
where the minus sign is conventional. 

Using the general solutions \eqref{gen-varepsilon} and \eqref{gen-chi} in the bulk expression \eqref{papaefgenrescrec3d}, and trading the chiral functions $\ell_{\pm}$ for $L_{\pm}$ defined as (the prime stands for the derivative with respect to the unique argument of the function)
 \begin{equation}
\ell_{\pm}=
\dfrac{1}{\left(\xi^{\pm}\right)^2}\left(L_{\pm}-\dfrac{(\xi^{\pm\prime})^2-2 \xi^{\pm}\xi^{\pm\prime\prime}}{4}\right),
\end{equation}
we obtain the following metric:
\begin{eqnarray}
\nonumber
\text{d}s^2_{\text{Einstein}} 
&=&-\frac{1}{k}
\left(\sqrt{-\frac{\xi^-}{\xi^+}}
\text{d}x^+ -
\sqrt{-\frac{\xi^+}{\xi^-}}
\text{d}x^-
\right)\text{d}r
\\
&&+\left(\frac{L_+}{k^2} 
-\frac{r}{2k}\sqrt{-\xi^+\xi^-}\xi^{+\prime}
\right) \left(\frac{\text{d}x^+}{\xi^+}\right)^2
+\left(\frac{L_-}{k^2} 
-\frac{r}{2k}\sqrt{-\xi^+\xi^-}\xi^{-\prime}
\right) \left(\frac{\text{d}x^-}{\xi^-}\right)^2
\nonumber
\\
&&+\left(
r^2 +\frac{r}{2k}\frac{1}{\sqrt{-\xi^+\xi^-}}
\left(
\xi^{+\prime}+
\xi^{-\prime}
\right)
+
\frac{L_++L_-}{k^2\xi^+ \xi^-} \right)\text{d}x^+\text{d}x^-.
\label{papaefgenrescrec3d-flb}
\end{eqnarray}
This metric depends on four arbitrary functions: $\xi^+(x^+)$ and $\xi^-(x^-)$ carrying  information about the holographic fluid velocity (see \eqref{flat-vel-comp}),  and  $L_+(x^+)$, $L_-(x^-)$, which together with  $\xi^+(x^+)$ and $\xi^-(x^-)$ shape the energy--momentum tensor -- here traceless due to the boundary flatness. Indeed we have
\begin{equation}
\label{flat-epschi}
\varepsilon= -\frac{1}{4\pi G }\frac{ L_++L_-}{\xi^+\xi^-},\quad
\chi=
\frac{1}{4\pi G }\frac{ L_+-L_-}{\xi^+\xi^-},
\end{equation}
and in turn
\begin{equation}
\label{empm}
T_{\pm\pm}=\frac{L_{\pm}}{4\pi G(\xi^{\pm})^2}.
\end{equation}

In three dimensions, any Einstein spacetime is locally anti-de Sitter. Hence, there exists always a coordinate transformation that can be used to bring it into a canonical $\text{AdS}_3$ form (say, in Poincar\'e coordinates). This is a large gauge transformation whenever the original Einstein spacetime has non-trivial conserved charges. The determination of the latter is therefore crucial for a faithful identification of the solution under consideration. It allows to evaluate the precise role played by the above arbitrary functions.

The charge computation requires a complete family of asymptotic Killing vectors. Those are determined according to the gauge, \emph{i.e.} to the fall-off behaviour at large-$r$. The family \eqref{papaefgenrescrec3d-flb} does not fit BMS gauge, unless $\xi^\pm$ are constant. This is equivalent to saying that the fluid has a uniform velocity, and can therefore be set at rest by an innocuous global Lorentz boost tuning  $\xi^+=1$ and $\xi^-=-1$.\footnote{Observe that one may reabsorb $\xi^+$ and $\xi^-$ by redefining $\text{d}x^\pm\to \xi^\pm\text{d}x^\pm$ and $r\to \nicefrac{r}{\sqrt{-\xi^+\xi^-}}$ inside \eqref{papaefgenrescrec3d-flb}. This does not prove, however, that  $\xi^\pm$ play no role, and this is why we treat them separately. }  We will first focus on this case, where the asymptotic Killing vectors are known, and move next to the other extreme, demanding the fluid be perfect, \emph{i.e.} in Landau--Lifshitz hydrodynamic frame. In the latter instance we will have to determine this family of vectors beforehand, as the gauge will no longer be BMS. Investigating the general situation captured by \eqref{papaefgenrescrec3d-flb} is not relevant for our argument, which is meant to show that fluid/gravity holographic reconstruction is hydrodynamic-frame dependent. 

As we will see, the charges computed following \cite{BarnichCG1, BarnichCG2, code},
and displayed in Eqs. \eqref{chmodesdis} and \eqref{chmodesperf}, coincide in both cases with the modes of the energy--momentum tensor \eqref{empm}. However, they obey a different algebra due to the distinct asymptotic behaviour of the associated metric families.

\paragraph{Dissipative static fluid}

As anticipated, this class of solutions is reached by demanding $\xi^\pm=\pm1$, while keeping $L^{\pm}$ arbitrary. We obtain
\begin{equation}
\label{Ein-dis}
\text{d}s^2_{\text{Einstein}} 
=-\frac{1}{k}
\left(
\text{d}  x^+ -
\text{d}  x^-
\right)\text{d}  r+  r^2 \text{d}  x^+\text{d}  x^-
+
\frac{1}{k^2} 
\left(L_+\text{d}  x^+
-
L_-\text{d}  x^-
\right)
 \left(\text{d}  x^+-\text{d}  x^-\right),
\end{equation}
which is the canonical expression of Ba\~nados solutions in BMS gauge. Following \eqref{flat-epschi}, the boundary fluid energy and heat densities are $\varepsilon= \nicefrac{1}{4\pi G }\left( L_++L_-\right)$ and $\chi = -\nicefrac{1}{4\pi G }\left( L_+-L_-\right)$. Therefore the heat current is not vanishing, and in the present hydrodynamic frame the fluid is at rest and dissipative. 

The class of metrics  \eqref{Ein-dis} are form-invariant under 
\begin{equation}
\label{Killzeta}
\zeta = \zeta^r \partial_r +   \zeta^+ \partial_+  + \zeta^- \partial_- 
\end{equation}
with
\beqn
\nonumber
\label{Killcompzetar}
\zeta^r&=& -\dfrac{r}{2}\left(Y^{+\prime}+Y^{-\prime}\right)+\dfrac{1}{2k}\left(Y^{+\prime\prime}-Y^{-\prime\prime}\right)
\\
&&-\dfrac{1}{2k^2 r}\left(L_{+}-L_{-}\right)\left(Y^{+\prime}-Y^{-\prime}\right),\\
\label{Killcompzetapm}
\zeta^\pm&=& Y^{\pm}-\dfrac{1}{2kr}\left(Y^{+\prime}-Y^{-\prime}\right),
\eeqn
for arbitrary chiral functions $Y^+(x^+)$ and   $Y^-(x^-)$.
These vector fields generate diffeomorphisms, which alter the functions  appearing in  \eqref{Ein-dis} 
according to 
\begin{equation}
\label{asympLie}
-\mathscr{L}_\zeta g_{MN}
=\delta_\zeta  g_{MN} 
= \frac{\partial g_{MN}}{\partial L_+}\delta_\zeta  L_+
+ \frac{\partial g_{MN}}{\partial L_-}\delta_\zeta  L_-
\end{equation}
with
\beq
\delta_{\zeta}{L_{\pm}}=-Y^{\pm}L_{\pm}^{\prime}-2L_{\pm}Y^{\pm\prime}+\frac{1}{2} Y^{\pm\prime\prime\prime}.
\eeq
The last term in this expression is responsible for the emergence of a central charge in the surface-charge algebra.  These vectors obey an algebra for the modified Lie bracket (see e.g. \cite{Barnich:2010eb}):
\begin{equation}
\label{modlie}
\zeta_3=\left[\zeta_1,\zeta_2\right]_{\text{M}}=\left[\zeta_1,\zeta_2\right]-\delta_{\zeta_2}\zeta_1+\delta_{\zeta_1}\zeta_2
\end{equation}
with\footnote{Here $\delta_{\zeta_2}\zeta_1$ stands for the variation produced on $\zeta_1$ by $\zeta_2$, and this is not vanishing because $\zeta_1$ depends explicitly on $L_{\pm}$:  $\delta_{\zeta_2}\zeta_1=
\left(\frac{\partial \zeta^N_1}{\partial L_+} 
\delta_{\zeta_2}L_+
+
\frac{\partial \zeta^N_1}{\partial L_-} 
\delta_{\zeta_2}L_-
\right)
\partial_N
$.}  $\zeta_a=\zeta\left(Y_a^+,Y_a^- \right)$ and
\begin{equation}
Y_3^\pm= Y_1^\pm Y_2^{\pm\prime}-Y_2^\pm Y_1^{\pm\prime}.
\end{equation}

The surface charges are computed for an arbitrary metric $g$ of the type \eqref{Ein-dis}  with global $\text{AdS}_3$ as reference background. The latter has metric $\bar{g}$ with $L_+=L_-=\nicefrac{-1}{4}$ \emph{i.e.} $\varepsilon=\nicefrac{-1}{8\pi G}$ and $\chi=0$.
The final integral is performed over the compact spatial boundary coordinate $x\in[0,2\pi]$:
\beq
Q_{Y}\left[g-\bar g,\bar{g}\right]=\dfrac{1}{8\pi k G}\int_0^{2\pi}\text{d}x \left(Y^+\left(L_++\frac{1}{4}\right)-Y^-\left(L_-+\frac{1}{4}\right)\right).
\eeq
These charges are in agreement with the quoted literature,\footnote{Some relative-sign differences are due to different conventions used for the light-cone coordinates, here defined as $x^{\pm}=x\pm kt$.} 
and their algebra is determined as usual:
\begin{equation}
\left\{Q_{Y_1}, 
Q_{Y_2}\right\}= \delta_{\zeta_1}Q_{Y_2}=-\delta_{\zeta_2}Q_{Y_1}.
\end{equation}
Introducing the modes
\begin{equation}
\label{chmodesdis}
L^\pm_m=\dfrac{1}{8\pi k G}\int_0^{2\pi}\text{d}x \, \text{e}^{\text{i}mx^\pm}\left(L_\pm+\frac{1}{4}\right)
\end{equation}
the algebra reads:
\begin{equation}
\text{i}\left\{L^\pm_m, L^\pm_n
\right\}=(m-n)L^\pm_{m+n}+\frac{c}{12}m\left(m^2-1\right)\delta_{m+n,0}\, , \quad
\left\{L^\pm_m, L^\mp_n
\right\}=0.
\end{equation}
This double realization of Virasoro algebra with Brown--Henneaux central charge $c=\nicefrac{3}{2kG}$ was expected for  Ba\~nados solutions \eqref{Ein-dis}. 

\paragraph{Perfect fluid with arbitrary velocity}

In Landau--Lifshitz frame the heat current vanishes ($\chi=0$) and the boundary conformal fluid is perfect. Equation  \eqref{flat-epschi} requires for this 
\beq
L_{+}=L_{-}=\dfrac{M}{2},
\eeq
with $M$ constant, while it gives for energy density $\varepsilon= -\nicefrac{M}{4\pi G \xi^+\xi^-}$.
As for the general case, the reconstructed bulk family of metrics 
\begin{eqnarray}
\nonumber
\text{d}s^2_{\text{Einstein}} 
&=&-\frac{1}{k}
\left(\sqrt{-\frac{\xi^-}{\xi^+}}
\text{d}x^+ -
\sqrt{-\frac{\xi^+}{\xi^-}}
\text{d}x^-
\right)\text{d}r
\\
&&+\left(\frac{M}{2k^2} 
-\frac{r}{2k}\sqrt{-\xi^+\xi^-}\xi^{+\prime}
\right) \left(\frac{\text{d}x^+}{\xi^+}\right)^2
+\left(\frac{M}{2k^2} 
-\frac{r}{2k}\sqrt{-\xi^+\xi^-}\xi^{-\prime}
\right) \left(\frac{\text{d}x^-}{\xi^-}\right)^2
\nonumber
\\
&&+\left(
r^2 +\frac{r}{2k}\frac{1}{\sqrt{-\xi^+\xi^-}}
\left(
\xi^{+\prime}+
\xi^{-\prime}
\right)
+
\frac{M}{k^2\xi^+ \xi^-} \right)\text{d}x^+\text{d}x^-
\label{papaefgenrescrec3d-perfect}
\end{eqnarray}
is not in BMS gauge, unless $\xi^\pm$ are constant. Again this latter subset is entirely captured by $\xi^\pm=\pm1$, and the resulting solution is BTZ together with all non-spinning zero-modes of Ba\~nados family \cite{BTZ1, BTZ2,banados3}:
\begin{equation}
\label{nospin}
\text{d}s^2_{\text{Einstein}} 
=-\frac{1}{k}
\left(
\text{d}  x^+ -
\text{d}  x^-
\right)\text{d}  r+  r^2 \text{d}  x^+\text{d}  x^-
+
\frac{M}{2k^2} 
\left(\text{d}  x^+
-
\text{d}  x^-
\right)^2
.
\end{equation}

The asymptotic structure rising in \eqref{papaefgenrescrec3d-perfect} is now respected by the following family of asymptotic Killing vectors
\begin{equation}
\label{adsKilletatild}
\eta = \eta^{r} \partial_{r} +   \eta^{+} \partial_{+}  + \eta^{-} \partial_{-},
\end{equation}
expressed in terms of two arbitrary chiral functions $\epsilon^\pm(x^\pm)$
\begin{equation}
\label{adsKilletatildcomp}
 \eta^{r} = -\frac{r}{2}\left(\epsilon^{+\prime} +\epsilon^{-\prime} \right),
\quad
 \eta^{\pm} = \epsilon^{\pm}.
\end{equation}
These vectors, slightly different from those found for the dissipative boundary fluids \eqref{Killzeta}, \eqref{Killcompzetar}, \eqref{Killcompzetapm}, appear as the result of an exhaustive analysis of \eqref{papaefgenrescrec3d-perfect}. They do not support  subleading terms, and since they do not depend on the the functions $\xi^\pm$, they form an algebra for the Lie bracket:
\begin{equation}
\left[\eta_1,\eta_2\right]=\eta_3 
\end{equation}
with
$\eta_a=\eta\left(\epsilon_a^+,\epsilon_a^- \right)$ and
\begin{equation}
  \epsilon_3^\pm= \epsilon_1^\pm \epsilon_2^{\pm\prime}-\epsilon_2^\pm \epsilon_1^{\pm\prime}.
\end{equation}
They induce the exact transformation 
\begin{equation}
\label{adsdelgzet}
-\mathscr{L}_\eta g_{MN}=\delta_\eta  g_{MN} 
=\frac{\partial g_{MN}}{\partial \xi^+ }\delta_\eta  \xi^+ 
+ \frac{\partial g_{MN}}{\partial \xi^{+\prime}}\delta_\eta  \xi^{+\prime}
+ \frac{\partial g_{MN}}{\partial \xi^-}\delta_\eta  \xi^-
+ \frac{\partial g_{MN}}{\partial \xi^{-\prime}}\delta_\eta  \xi^{-\prime}
\end{equation}
 with 
\begin{equation}
 \delta_\eta  \xi^\pm =\xi^\pm \epsilon^{\pm\prime} -  \epsilon^\pm \xi^{\pm\prime}. 
 \end{equation}

Following the customary pattern, we can determine the conserved charges, with global $\text{AdS}_3$ as reference background, now reached with $\xi^\pm=\pm1$ and $M=\nicefrac{-1}{2}$ (again $\varepsilon=\nicefrac{-1}{8\pi G}$ and $\chi=0$):
\beq
Q_{\epsilon}\left[g-\bar g,\bar{g}\right]=\dfrac{1}{16\pi k G}\int_0^{2\pi}\text{d}x\left(\epsilon^+\left(\frac{1}{\xi^{+2}}-1\right)-\epsilon^-\left(\frac{1}{\xi^{-2}}-1\right)\right),
\eeq
as well as their algebra:
\begin{equation}
\left\{Q_{\epsilon_1}, 
Q_{\epsilon_2}\right\}= \delta_{\eta_1}Q_{\epsilon_2}=-\delta_{\eta_2}Q_{\epsilon_1}.
\end{equation}
Defining now
\begin{equation}
\label{chmodesperf}
Z^\pm_m=\dfrac{1}{16\pi k G}\int_0^{2\pi}\text{d}x \, \text{e}^{\text{i}mx^\pm}\left(\frac{1}{\xi^{\pm2}}-1\right)
\end{equation}
we find
\begin{equation}
\label{deWitt}
\text{i}\left\{Z^\pm_m, Z^\pm_n
\right\}=(m-n)Z^\pm_{m+n}+\frac{m}{4kG}\delta_{m+n,0}\, , \quad
\left\{Z^\pm_m, Z^\mp_n
\right\}=0.
\end{equation}
The central extension of this algebra is trivial. Indeed, it can be reabsorbed in the following redefinition of the modes $Z^{\pm}_m$
\beq
\tilde Z^{\pm}_m=Z^{\pm}_m+\dfrac{1}{8k G}\delta_{m,0}.
\eeq
Therefore, \eqref{deWitt} becomes
\beq
\label{deWitt1}
\text{i}\left\{\tilde Z^\pm_m, \tilde Z^\pm_n
\right\}=(m-n)\tilde Z^\pm_{m+n}, \quad
\left\{\tilde Z^\pm_m, \tilde Z^\mp_n
\right\}=0.
\eeq
The algebra at hand \eqref{deWitt1} is de Witt rather than Virasoro,\footnote{The absence of central charges occurs also in \cite{Troessaert:2013fma} for the same reason, \emph{i.e.} a modification of the asymptotic behaviour.} and this outcome demonstrates the already advertised result: the family of locally anti-de Sitter spacetimes obtained holographically from two-dimensional fluids in the Landau--Lifshitz frame overlap only partially the space of Ba\~nados solutions.  This overlap encompasses the non-spinning BTZ and excess or defects geometries provided in \eqref{nospin}.

\subsection{Charges in Ricci-flat spacetimes}\label{sec:recon-flat-bry-flat}

The absence of anomaly in the Carrollian framework is equivalent to setting $\Sigma^{x}_{\hphantom{x}x}=\Xi^{x}_{\hphantom{x}x}=0$ (see \eqref{Xiexpchi-anomaly}), whereas the Weyl--Carroll flatness requires $s=0$ (see \eqref{dyn-WCc}). This amounts to taking $\Omega=a=1$ and $b_x=0$,\footnote{Actually the absence of anomaly requires rather $\Omega=\Omega(t)$, $a=a(x)$ and $b_x=b_x(x)$, which can be reabsorbed trivially with Carrollian diffeomorphisms.} and with those data 
$s=0$ reads
\begin{equation}
\label{carcond-expl-WCflat-sol}
\partial_t^2\beta_x=0.
\end{equation}
In the Carrollian spacetime at hand, the fluid equations of motion
\eqref{flat3dfinbetasimple} are
 \begin{equation}
 \label{flat3dfin-ex3} 
\begin{cases}
&\partial_t\varepsilon=0,
\\
&\partial_x \varepsilon
+\partial_t (\pi_x+2\varepsilon \beta_{x})
= 0
.
\end{cases}
\end{equation}

Equations \eqref{carcond-expl-WCflat-sol} and \eqref{flat3dfin-ex3} can be integrated in terms of four arbitrary functions of $x$: $\varepsilon(x)$, $\varpi(x)$, $\lambda(x)$ and $\mu(x)$. We find
\begin{eqnarray}
\label{carcond-expl-flat-sol}
\beta_x(t,x)&=& \frac{\lambda(x)}{2\varepsilon(x)}-\frac{t}{2}\partial_x\ln\mu(x)
,\\
 \label{pigen}
\pi_{x}(t,x)&=&-2\varepsilon(x) \beta_{x}(t,x)
+\varpi(x)-t \partial_x\varepsilon(x)
\end{eqnarray}
(this parameterization of $\beta_x$ will be appreciated later). The Ricci-flat (even locally flat) holographically reconstructed spacetime from these Carrollian fluid data is obtained from the general expression \eqref{papaefresricf}:
\begin{equation}
\begin{split}
\text{d}s^2_{\text{flat}} =&
-2\left(\text{d}t - \beta_x \text{d}x\right)\left( \text{d}r+r \partial_t \beta_x \text{d}x\right)+r^2\text{d}x^2
 \\
 &+8\pi G \left(\varepsilon(\text{d}t- \beta_x \text{d}x)^2
 -\pi_x \text{d}x (\text{d}t-\beta_x\text{d}x)\right),
 \end{split}
\label{papaefresricf-ex-3}
\end{equation}
where $\beta_x $ and $\pi_x$ are meant to be as in \eqref{carcond-expl-flat-sol} and \eqref{pigen}.

On the one hand, the arbitrary functions 
$ \varepsilon(x)$ and $\varpi(x)$ 
are reminiscent of the functions 
$L_\pm(x^\pm)$ (or $ \varepsilon(t,x)$ and $\chi(t,x)$)
present in the AdS solutions. 
A vanishing-$k$ limit was indeed used in Ref. \cite{Glenn3} to obtain $\varepsilon(x)$ and $\varpi(x)$ from $L_\pm(x^\pm)$. On the other hand, $\lambda(x)$ and $\mu(x)$ 
remind $\xi^\pm(x^\pm)$, and are indeed a manifestation of a residual hydrodynamic frame invariance, which survives the Carrollian limit. Considering indeed the Carrollian hydrodynamic-frame transformations \eqref{betacompocar}
\begin{equation}
\beta_x^\prime
=\beta_x+B_x,
\end{equation}
in the present framework ($\Sigma^{x}_{\hphantom{x}x}=\Xi^{x}_{\hphantom{x}x}=0$), and using Eqs. \eqref{vare-trans}, \eqref{chi-trans}, \eqref{qchiprime}, \eqref{tautauprime}, \eqref{QexpC}, \eqref{chidec}, \eqref{Qexpchi}, we obtain the transformations:
\begin{equation}
\varepsilon^\prime=\varepsilon,\quad 
\pi_x^\prime=\pi_x-2\varepsilon B_x,
\end{equation}
which leave the Carrollian fluid  equations \eqref{flat3dfin-ex3} invariant.  
The new velocity field $\beta_x^\prime$ is compatible with the Weyl--Carroll flatness \eqref{carcond-expl-WCflat-sol} provided the transformation function $B_x$ is linear in time, hence parameterized in terms of two arbitrary functions of $x$. This is how $\lambda(x)$ and $\mu(x)$ emerge. 

Observe also that the residual Carrollian hydrodynamic frame invariance enables us to define here a Carrollian Landau--Lifshitz hydrodynamic frame. Indeed, combining \eqref{pigen} and \eqref{carcond-expl-flat-sol} we obtain
\begin{equation}
 \label{pigen2}
\pi_{x}(t,x)=-\lambda(x)
+\varpi(x)+t \varepsilon(x)\partial_x\ln \frac{\mu(x)}{\varepsilon(x)}.
\end{equation}
Adjusting the velocity field $\beta_x$ such that
\begin{equation}
\label{perfl-car}
\varpi(x)=\lambda(x) \quad \text{and}\quad \frac{\varepsilon(x)}{\mu(x)}={\varepsilon_0}
\end{equation}
with $\varepsilon_0$ a constant, makes the Carrollian fluid perfect: $\pi_x=0$.

In complete analogy with the AdS analysis, we will first compute the charges for vanishing velocity $\beta_x=0$ (which is given by $\lambda(x)=0$ and $\mu(x)=1$) in terms of $\varepsilon(x)$ and $\varpi(x)$, and next perform the similar computation for  perfect fluids with velocity $\beta_x$ parameterized with two arbitrary functions $\lambda(x)$ and $\mu(x)$. Here empty Minkowski bulk is realized with   $\mu=1$, $\lambda=0$, $\varpi=0$ and $\varepsilon_0 =\nicefrac{-1}{8\pi G}$.

As for the AdS instance discussed in Sec. \ref{sec:recon-flat-bry-AdS}, the class \eqref{papaefresricf-ex-3} is not in the BMS gauge, unless $\beta_x$ is constant, which can then be reabsorbed by a global Carrollian boost (constant $B_x$).\footnote{The functions 
$\lambda(x)$ and $\mu(x)$ entering \eqref{papaefresricf-ex-3}  via \eqref{pigen} and \eqref{carcond-expl-flat-sol} can be reabsorbed in any case by performing the coordinate transformation $\text{d}x\to \frac{\text{d}x}{\sqrt{\mu(x)}}$, $\text{d} t \to \frac{1}{\sqrt{\mu(x)}}  \left(\text{d} t+\beta_x
\text{d}x\right)$
and
$r\to {r}{\sqrt{\mu(x)}}$. This leads to the same form as the one reached by setting $\mu=1$ and $\lambda=0$,  \emph{i.e}
 \eqref{papaefresricf-ex-3-flat}. 
\label{footchange}} We will first discuss this situation, where the asymptotic Killings are the canonical generators of $\mathfrak{bms}_3$. Outside the BMS, we will determine the asymptotic isometry for metrics reconstructed from perfect fluids, and proceed with the surface charges and their algebra. Our conclusion is here that asymptotically flat fluid/gravity correspondence is sensitive to the residual  hydrodynamic-frame invariance.  

\paragraph{Dissipative static fluid}
  
The metric \eqref{papaefresricf-ex-3} for vanishing $\beta_x$ takes the simple form (again the prime signals a derivative)
\begin{equation}
\text{d}s^2_{\text{flat}} =
-2\text{d} t \text{d} r+ r^2\text{d} x^2
 +8\pi G \left(  \varepsilon \text{d} t 
 - \left( \varpi - t \varepsilon^{\prime}\right) \text{d} x 
 \right) \text{d} t,
 \label{papaefresricf-ex-3-flat}
\end{equation}
compatible with BMS gauge with asymptotic Killing vectors
\beq
\zeta=\zeta^r \pa_r+\zeta^t \pa_t+\zeta^x \pa_x,
\eeq
where
\beqn
\zeta^r &=& -r Y^{\prime}+H^{\prime\prime}+t Y^{\prime\prime\prime}+\dfrac{4\pi G}{r}\left(\varpi-t\varepsilon^{\prime}\right)\left(H^{\prime}+t Y^{\prime\prime}\right),\label{zetaflatr}\\
\zeta^t &=& H+t Y^{\prime},  \label{zetaflatt}\\
\zeta^x &=& Y-\dfrac{1}{r}\left(H^{\prime}+t Y^{\prime\prime}\right).\label{zetaflatx}
\eeqn
Here $H$ and $Y$ are functions of $x$ only. Vectors \eqref{zetaflatr},  \eqref{zetaflatt},  \eqref{zetaflatx} are the vanishing-$k$ limit of \eqref{Killzeta}, \eqref{Killcompzetar}, \eqref{Killcompzetapm}, reached by trading light-cone frame as $x^\pm=x\pm kt $, and setting $Y^{\pm}(x^\pm)=Y(x)\pm k\left(H(x)+t Y^{\prime}(x)\right)$. 

 This family of vectors produces the following variation on the metric fields:
\beq
-\mathscr{L}_\zeta g_{ MN}=\delta_\zeta g_{ MN}
=\frac{\partial g_{ MN}}{\partial \varepsilon }\delta_\zeta \varepsilon +\frac{\partial g_{ MN}}{\partial\varepsilon^{\prime}}\delta_\zeta \varepsilon^{\prime}
+ \frac{\partial g_{ MN}}{\partial \varpi}\delta_\zeta  \varpi,
\eeq
with
\beqn
\delta_{\zeta}\varepsilon &=& -2\varepsilon Y^{\prime}-Y  \varepsilon^{\prime}+\frac{Y^{\prime\prime\prime}}{4\pi G},\\
\delta_{\zeta}\varpi &=&-\dfrac{H^{\prime\prime\prime}}{4\pi G}+\dfrac{1}{H}\left(\varepsilon H^2\right)^{\prime}-\dfrac{1}{Y} \left(\varpi Y^2\right)^{\prime}.
\eeqn
Their algebra closes for the same modified Lie bracket \eqref{modlie} with $\zeta_a=\zeta\left(H_a,Y_a \right)$ and
\begin{equation}
Y_3=Y_1 Y_2^{\prime}-Y_2 Y_1^{\prime}
\quad
H_3=Y_1H_2^{\prime}+H_1Y_2^{\prime}-Y_2H_1^{\prime}-H_2Y_1^{\prime}.
 \end{equation}

We can compute the charges of $g$ in \eqref{papaefresricf-ex-3-flat}, using Minkowski as reference background $\bar{g}$. They read:
\begin{equation}
Q_{H,Y}[g-\bar{g},\bar{g}]=\frac12\int_0^{2\pi}\text{d} x \left[
H\left( \varepsilon+\frac{1}{8\pi G}\right)-Y  \varpi
\right].
\end{equation}
With a basis of functions $\exp \text{i} m x$ for $H$ and $Y$, we find the standard collection of charges
\begin{equation}
\label{chmodesdisC}
P_m=\frac12\int_0^{2\pi}\text{d} x \, 
\text{e}^{\text{i} mx}\left( \varepsilon+\frac{1}{8\pi G}\right), 
\quad J_m=-\frac12\int_0^{2\pi}\text{d} x  \, 
\text{e}^{\text{i} mx}  \varpi,
\end{equation}
which coincide with the computation performed \emph{e.g.} in \cite{Glenn3}. Using 
\begin{equation}
\left\{Q_{H_1,Y_1}, 
Q_{H_2,Y_2}\right\}= \delta_{\zeta_1}Q_{H_2,Y_2}=-\delta_{\zeta_2}Q_{H_1,Y_1},
\end{equation}
we obtain the following surface-charge algebra:
\begin{equation}
\text{i}\left\{J_m, P_n
\right\}=(m-n)P_{m+n}+\frac{c}{12}m\left(m^2-1\right)\delta_{m+n,0}\, ,
\quad
\text{i}\left\{J_m, J_n
\right\}=(m-n)J_{m+n}
\, , \quad
\left\{P_m, P_n
\right\}=0
\end{equation}
with $c=\nicefrac{3}{G}$. This is the $\mathfrak{bms}_3$ algebra, and this analysis 
demonstrates that a non-perfect Carrollian fluid, even with $\beta_x=0$, is sufficient for generating holographically all Barnich--Troessaert flat three-dimensional spacetimes. This goes along with the analogue conclusion reached in AdS for Ba\~nados spacetimes.  

\paragraph{Perfect fluid with velocity}

Consider now the resummed metric \eqref{papaefresricf-ex-3} assuming \eqref{perfl-car}. We obtain
\begin{equation}
\text{d}s^2_{\text{flat}} =
-2\left(\text{d}t - \beta_x \text{d}x\right)\left( \text{d}r-  \frac{r\mu^\prime}{2\mu} \text{d}x\right)+r^2\text{d}x^2
 +8\pi G  \varepsilon_0 \mu \left(\text{d}t - \beta_x \text{d}x\right) ^2
\label{papaefresricf-ex-2}
\end{equation}
with $\beta_x $ given by
\begin{equation}
\beta_x=\frac{1}{2\mu}\left(\frac{\lambda}{\varepsilon_0}-t \mu^{\prime}\right).
\end{equation}
Unless $\beta_x$ is constant, the metrics \eqref{papaefresricf-ex-2}
are not in BMS gauge. The BMS subset is entirely captured by $\mu=1$, $\lambda=0$ with resulting solutions plain Minkowski ($\varepsilon_0 =\nicefrac{-1}{8\pi G}$)  and the non-spinning zero-modes of Barnich--Troessaert family:
\begin{equation}
\text{d}s^2_{\text{flat}} = -2\text{d}  t \text{d}  r+  r^2 \text{d}  x^2+8\pi G  \varepsilon_0 \text{d}  t^2. \label{mink1}
\end{equation}

The asymptotic isometries of \eqref{papaefresricf-ex-2} are now generated by\footnote{Again the fields \eqref{Killetatild}, \eqref{Killetatildcomp} are alternatively obtained by an appropriate zero-$k$ limit of \eqref{adsKilletatild} and \eqref{adsKilletatildcomp}.} 
\begin{equation}
\label{Killetatild}
\eta = \eta^{r} \partial_{r} +   \eta^{t} \partial_{t}  + \eta^{x} \partial_{x},
\end{equation}
expressed in terms of two arbitrary functions $h({x})$ and $\rho({x})$
\begin{equation}
\label{Killetatildcomp}
\eta^{r} = -r\rho^{\prime}, \quad
\eta^{t} =  h +t \rho^{\prime}, \quad
\eta^{x} = \rho.
\end{equation}
The algebra of asymptotic Killing vectors closes for the ordinary Lie bracket 
\begin{equation}
\left[\eta_1,\eta_2\right]=\eta_3
\end{equation}
with $\eta_a=\eta\left(h_a,\rho_a \right)$ and 
\begin{equation}
\rho_3= \rho_1 \rho_2^{\prime}-\rho_2 \rho_1^{\prime}, \quad h_3=\rho_1  h_2^{\prime}+h_1 \rho_2^{\prime}-\rho_2  h_1^{\prime}-h_2 \rho_1^{\prime}.
\end{equation}
It respects the form of the metric
\begin{equation}
\label{delgzet}
-\mathscr{L}_\eta g_{ MN}=\delta_\eta g_{ MN}
=\frac{\partial g_{ MN}}{\partial \mu }\delta_\eta  \mu 
+ \frac{\partial g_{ MN}}{\partial \mu^{\prime}}\delta_\eta  \mu^{\prime}
+ \frac{\partial g_{ MN}}{\partial \lambda}\delta_\eta  \lambda
\end{equation}
 with 
\begin{eqnarray}
\delta_\eta \lambda &=&-2 \lambda \rho^{\prime}-\rho \lambda^{\prime}+ \varepsilon_0 \left(2\mu h^{\prime} +h\mu^{\prime}\right),\\
\delta_\eta \mu &=&-2\mu\rho^{\prime}-\rho\mu^{\prime}.
 \label{delta}
\end{eqnarray}

The charges of $g$ in \eqref{papaefresricf-ex-2} are computed as usual with Minkowski as reference background $\bar{g}$. They read:
\begin{equation}
Q_{h,\rho}[g-\bar{g},\bar{g}]=\frac12\int_0^{2\pi}\text{d} x \left[
h\left(\varepsilon_0\mu+\frac{1}{8\pi G}\right)-\rho  \lambda
\right].
\end{equation}
With a basis of unimodular exponentials for $h$ and $\rho$, we find now
\begin{equation}
\label{chmodesperfC}
M_m=\frac12\int_0^{2\pi}\text{d} x \, 
\text{e}^{\text{i} mx}\left( \varepsilon_0\mu+\frac{1}{8\pi G}\right), 
\quad I_m=-\frac12\int_0^{2\pi}\text{d} x  \, 
\text{e}^{\text{i} mx}  \lambda,
\end{equation}
and 
\begin{equation}
\left\{Q_{h_1,\rho_1}, 
Q_{h_2,\rho_2}\right\}= \delta_{\eta_1}Q_{h_2,\rho_2}=-\delta_{\eta_2}Q_{h_1,\rho_1}
\end{equation}
provide the surface-charge algebra:
\begin{equation}
\text{i}\left\{I_m, M_n
\right\}=(m-n)M_{m+n}-\frac{m}{4G}\delta_{m+n,0}\, , \quad\text{i}\left\{I_m, I_n
\right\}=(m-n)I_{m+n}\, , \quad
\left\{M_m, M_n
\right\}=0.
\end{equation}
As for the anti-de Sitter case, the central extension of this algebra is trivial. By translating the modes
\beq
\tilde M_m=M_m-\dfrac{1}{8G}\delta_{m,0},
\eeq
we obtain
\begin{equation}
\text{i}\left\{I_m, \tilde M_n
\right\}=(m-n)\tilde M_{m+n}, \quad\text{i}\left\{I_m, I_n
\right\}=(m-n)I_{m+n}\, , \quad
\left\{\tilde M_m, \tilde M_n
\right\}=0.
\end{equation}

This  algebra, which could have been obtained from \eqref{deWitt1} in the zero-$k$ limit, has no central charge. Therefore, our computation shows unquestionably that holographic locally flat spacetimes based on perfect Carrollian fluids -- fluids in Carrollian Landau--Lifshitz frame -- cover only in some measure the family on Barnich--Troessaert solutions. Among those one finds \eqref{mink1}.

\section{Conclusion}\label{se:conc}
We can now summarize our achievements. The motivations of the present work have been twofold: (\romannumeral1) reconstruct asymptotically anti-de Sitter and flat three-dimensional spacetimes using fluid/gravity holographic correspondence in a unified framework;  (\romannumeral2) investigate the emergence of  hydrodynamic-frame invariance and its potential holographic breakdown.

Solutions to three-dimensional vacuum Einstein's equations have been searched systematically since the seminal work of BTZ, and their asymptotic symmetries as well as the corresponding conserved charges are thoroughly understood. In parallel, many aspects of their boundary properties in the anti-de Sitter case were discussed before the advent of the holographic correspondence, and lately for the flat case in relation with the BMS asymptotic symmetries. However, setting up a precise correspondence between a general two-dimensional relativistic fluid defined on an arbitrary background and a three-dimensional  anti-de Sitter spacetime was only superficially analyzed, whereas the possible relationship among flat spacetimes and Carrollian fluid dynamics had never been considered. This has been the core of our inquiry. 

Because relativistic fluid dynamics in two spacetime dimensions is rather simple, it allows to perform an exhaustive and exact study of the equations of motion, and of their form invariance under hydrodynamic-frame transformations -- local Lorentz boosts.  We have assumed for commodity a conformal equation of state, keeping the fluid non-conformal though (\emph{i.e.} with non-zero viscous bulk pressure). Hence, the relativistic fluid is described by an arbitrary velocity field, the energy and heat densities, and the viscous pressure, all transforming appropriately under local Lorentz boosts so as to keep the energy--momentum tensor invariant. The extreme situation corresponds to the  Landau--Lifshitz frame, where the heat current vanishes and the energy--momentum tensor is diagonal. 

Three-dimensional Einstein spacetime reconstruction is then achieved with the derivative expansion,  following the usual pattern of higher dimensions. Here it is not an expansion but a finite sum, involving all boundary data. Holographic fluids have an anomalous viscous pressure proportional to the curvature of the host geometry. Owing to this fact, the holographic fluid does not move freely, but is subject to a force, entirely determined by its kinematical configuration and by the geometry.
Using light-cone coordinates and conformally flat boundary makes it easy to obtain the general fluid configuration, and a general and closed expression for locally anti-de Sitter spacetimes, in a gauge which is less stringent than BMS.

With this general result, it is possible to address the question of whether a boundary fluid configuration observed from different hydrodynamic frames gives rise to distinct bulk geometries. This is discussed in the simpler (but sufficient for the argument) case of flat boundaries with vanishing Weyl curvature, for which the fluid is conformal (no trace). The reconstructed bulk geometries are then described in terms of two pairs of chiral functions, $\xi^\pm$ and $L_\pm$. The former parameterize the velocity of the fluid, while the latter its energy and heat densities. With these data two extreme configurations emerge: (\romannumeral1) a fluid at rest with heat current; (\romannumeral2) a fluid with arbitrary velocity and vanishing heat current (hence perfect since the viscous pressure is also zero) \emph{i.e.}  in the Landau--Lifshitz frame. For both  cases one determines the bulk asymptotic Killing vectors together with the algebra of conserved surface charges. In the first instance, the left and right Virasoro algebras appear with their canonical central charges. In the second, the central charges can be reabsorbed by a redefinition of the elementary modes, demonstrating thereby that the bulk-metric derivative expansion is sensitive to the boundary-fluid hydrodynamic frame. In particular, the Landau--Lifshitz frame fails to reproduce faithfully all Ba\~nados' solutions, contrary to the common expectation. 

The above pattern has been resumed for the Ricci-flat spacetimes. The conformal boundary is now at null infinity, and is endowed with a Carrollian $1+1$-dimensional structure. Boundary dynamics is carried by a Carrollian fluid, obeying a set of hydrodynamic equations for energy and heat densities, two viscous stress scalars as well as a kinematic variable referred to as ``inverse-velocity''. Generically, these equations do not exhibit any sort of hydrodynamic-frame invariance. 

The reconstruction of three-dimensional Ricci-flat spacetimes is achieved by considering the vanishing-$k$ limit of the anti-de Sitter derivative expansion, which is finite. Information is supplied in this Ricci-flat derivative expansion by the Carrollian fluid defined at null infinity.  In particular, the original conformal anomaly is carefully identified as a source of Carrollian stress. 

As for Einstein spacetimes, we do not consider the most general situation, but impose equivalent restrictions: absence of anomaly and zero Weyl--Carroll curvature. The derivative-expansion gauge is slightly less restrained than BMS, and a residual hydrodynamic-frame-like invariance emerges, which allows to treat the same Carrollian dynamics from two equivalent perspectives: (\romannumeral1) a Carrollian fluid with vanishing inverse velocity and non-zero heat current; (\romannumeral2) a Carrollian fluid with inverse velocity and vanishing heat current (\emph{i.e.} a sort of Carrollian Landau--Lifshitz frame). Although equivalent from the Carrollian-fluid perspective, these two patterns lead to Ricci-flat spacetimes with different surface charge algebras. The former family fits in BMS gauge and reproduces all Barnich--Troessaert spacetimes with the appropriate charges. The  algebra is $\mathfrak{bms}_3$ with central charge. The set of Ricci-flat metrics obtained with a Carrollian perfect fluids exhibit an algebra whose central charge can be ultimately reabsorbed. 

The above is the bottom line of our work. Our findings raise several questions that we briefly sort in the following as possible physical applications, in three dimensions or beyond, and on either side of the fluid/gravity holographic correspondence.

At the first place, it is legitimate to ask where the origin of the hydrodynamic-frame invariance breaking stands.  We have implicitly or explicitly stated in our presentation that the responsible agent was fluid/gravity duality. This view is supported by the explicit expressions of surface charges (Eqs. \eqref{chmodesdis}, \eqref{chmodesperf}, \eqref{chmodesdisC} and \eqref{chmodesperfC}),  which appear as modes of the energy--momentum tensor for the relativistic fluid (or its Carrollian descendants), irrespective of the chosen velocity field. The breaking then occurs in the structure of the algebra, which is sensitive to the bulk-metric asymptotic behaviour, itself depending on the boundary-fluid velocity congruence. This reasoning is not a proof, and does not exclude that relativistic fluids might be, in their own right, globally sensitive to the locally arbitrary velocity field.\footnote{Changing hydrodynamic frame is a gauge transformation. As such, it can affect global properties.} 
Furthermore, our discussion has been confined to three bulk dimensions, where the observed breaking is necessarily global, as opposed to local (in three dimensions asymptotically AdS or flat translates into locally AdS and Minkowskian). Nothing excludes \emph{a priori} that in higher dimension, other obstructions of purely local nature emerge against the free choice of a relativistic congruence. The possible breakdown of the Landau--Lifshitz-frame paradigm has been quoted indeed for three-dimensional fluids in \cite{Ciambelli:2017wou}, in relation with the entropy current. No general concrete results are available at present though, and these questions remain relevant both for fluid dynamics and for the subject of fluid/gravity correspondence.

The second important issue concerns the systematic analysis of asymptotic Killing vectors and conserved charges for the fall-offs suggested by the derivative expansion. This question is valid in both anti-de Sitter
(Eq. \eqref{papaefgenrescrec3d}, or the further restricted versions presented in Sec. \ref{sec:reconAdS}) 
and flat spacetime 
(Eq. \eqref{papaefresricf} 
and other realizations in Sec. \ref{sec:reconflat}). In this respect, one should remind that the investigation of fall-off conditions generalizing Brown--Henneaux's was carried in Refs. \cite{Troessaert:2013fma,Perez:2016vqo, Grumiller:2016pqb, Grumiller:2017sjh}. Finding solutions to Einstein's equations obeying these more general asymptotic behaviours, \emph{i.e.} standing beyond Ba\~nados or Barnich--Troessaert, persists, and is worth pursuing in our framework (see the comment after Eq. 
 \eqref{gen-chi} and Ref. \cite{oblak}). In parallel, the Ricci-flat case calls for a deeper Hamiltonian understanding of the charges within the appropriate intrinsic Carrollian setup recently developed in \cite{CM1}. 

This latter comment opens Pandora's box for Carrollian physics, \emph{i.e.} physics in the ultra-relativistic regime, which is generally unexplored in a systematic fashion. Our study of Sec. \ref{sec:carfluid}, and Eqs. \eqref{carE}--\eqref{carH} in particular, exhibit the dynamics of two-dimensional ultra-relativistic fluids. It is remarkable that these physical systems are dual to Ricci-flat spacetimes. Equation  \eqref{papaefgenrescrec3d} is instrumental in setting this duality: it starts from the ordinary relativistic regime and reaches the Carrollian limit, from the gravitational side, as a Ricci-flat limit. This formalism is expected to have genuine physical applications in many-body one-dimensional systems -- and beyond one space dimension, as discussed in \cite{CMPPS2}.

Last and aside from the interplay between gravity and fluids, a purely hydrodynamic issue was also discussed, which remains puzzling: the entropy current. No microscopic definition or closed expression exist and this object is usually constructed order-by-order in the derivative expansion, physically restricted to comply with fundamental laws. In relativistic systems, this current is expected to be hydrodynamic-frame invariant, by essence of this invariance. Hence, any obstruction to the existence of such a frame-invariant current might dispute or hamper the freedom of choosing at wish the fluid velocity field. In two dimensions, we have the possibility to implement frame invariance exactly and we proposed a closed expression, which however is not unique and deserves further investigation. One should understand whether and why this is the proper choice, and possibly wonder if it provides a helpful guideline for handling the entropy current in systems of dimension higher than two.  Ultimately, in the spirit of considering its Carrollian limit, one should try to give a meaning to entropy in ultra-relativistic systems.

\section*{Acknowledgements}

We would like to acknowledge valuable scientific exchanges with G. Barnich, G. Bossard, S.~Detournay, D. Grumiller, M. Haack, R. Leigh, O. Miskovic, B. Oblak, R. Olea, T. Petkou and C. Zwikel. We also thank  A. Sagnotti for the \textsl{Workshop on Future of Fundamental Physics} (within the  6th International Conference on New Frontiers in Physics -- ICNFP), Kolybari, Greece, August 2017, where this work was initiated, as well as  E. Bergshoeff, N.~Obers and D. T. Son for the \textsl{Workshop on Applied Newton--Cartan Geometry} held in Mainz, Germany, March 2018, for a stimulating and fruitful atmosphere, where these ideas were further developed. We thank each others home institutions for hospitality and financial support. This work was partly funded by the ANR-16-CE31-0004 contract \textsl{Black-dS-String}.

\appendix

\end{document}